\documentstyle[12pt]{article}
\begin{document}
\title{Admissible states in quantum phase space}
\author{Nuno Costa Dias\footnote{{\it ncdias@mail.telepac.pt}} \\ Jo\~{a}o Nuno Prata\footnote{{\it joao.prata@ulusofona.pt}} \\ {\it Departamento de Matem\'atica} \\
{\it Universidade Lus\'ofona de Humanidades e Tecnologias} \\ {\it Av. Campo Grande, 376, 1749-024 Lisboa, Portugal}}

\maketitle
\begin{abstract}
We address the question of which phase space functionals might represent a quantum state. We derive necessary and sufficient conditions for both pure and mixed phase space quantum states. From the pure state quantum condition we obtain a formula for the momentum correlations of arbitrary order and derive explicit expressions for the wavefunctions in terms of time dependent and independent Wigner functions. We show that the pure state quantum condition is preserved by the Moyal (but not by the classical Liouville) time evolution and is consistent with a generic stargenvalue equation. As a by-product Baker's converse construction is generalized both to an arbitrary stargenvalue equation, associated to a generic phase space symbol, as well as to the time dependent case. These results are properly extended to the mixed state quantum condition, which is proved to imply the Heisenberg uncertainty relations. Globally, this formalism yields the complete characterization of the kinematical structure of Wigner quantum mechanics. The previous results are then succinctly generalized for various quasi-distributions. 
Finally, the formalism is illustrated through the simple examples of the harmonic oscillator and the free Gaussian wave packet. As a by-product, we obtain in the former example an integral representation of the Hermite polynomials. 
  
\end{abstract}

\section{Introduction}

The Weyl-Wigner approach to quantum mechanics \cite{Lee}-\cite{Dias2} constitutes an alternative consistent formulation of quantum theory. Its main advantage resides in the fact that both observables and states are represented by ordinary c-numbers in a non-commutative phase space and that this representation strongly mimics classical statistical mechanics. To any given quantum operator $\hat A$ from the quantum algebra $\hat{\cal A}$ we can attribute a "classical" phase space counterpart $A$ in the classical algebra ${\cal A}$ through Weyl's correspondence rule \cite{Weyl}. Likewise, there is a phase space entity called the Wigner function $F^W (x,p)$ \cite{Wigner}, which plays the role of the density matrix. Weyl's correspondence rule endows the phase space with the structures of a non-commutative product and a bracket (the Moyal bracket). All the usual quantities of interest in operator quantum mechanics (probabilities, mean values) can be evaluated in phase space through simple formulae reminiscent of classical statistical mechanics. 

The intuitive methods of deformation quantization have found interesting applications in a wide range of fields of research. Standard applications in non-relativistic quantum mechanics include the fields of the semiclassical limit of quantum mechanics \cite{Lee,Flato,Lee2,Smith,nuno3}, hybrid dynamics \cite{nuno2}, collision theory \cite{Lee,Carruthers} and quantum chaos \cite{Lee,latka,shin,Takahashi}. The deformation techniques have also been used to provide an alternative phase space formulation of quantum field theory \cite{Zachos1,Antonsen1} and to address the quantization of constrained dynamical systems \cite{Antonsen2}. Moreover, the deformed algebraic structures play a key part in some current developments in $M$-theory \cite{Fairlie4}-\cite{Pinzul}. The vast number of applications reinforces the general importance of the Wigner theory not only as an alternative formulation of quantum mechanics but also as a comparatively simpler formalism where to test and further develop the deformation approach. 

A problem lying at the core of Wigner theory is whether its formulation is fully consistent with standard operator quantum mechanics. The logical equivalence of the two formulations presupposes that we should be able to formulate quantum mechanics in phase space autonomously and subsequently be able to rederive operator quantum mechanics.
This program has had a long history. In refs.\cite{Weyl}-\cite{Flato} the phase space formulation of quantum mechanics was firmly established. In refs.\cite{Dias6} and \cite{Baker}-\cite{Tatarskii} the inverse route was explored: starting from phase space one tried to recover the operator formulation. In our opinion this path has not been fully completed.

In this work we wish to reinforce the general belief that the phase space formulation of quantum mechanics is a self-consistent autonomous theory in its own right by addressing the following issues: (i) given some real, normalizable phase space function $F^W (x,p)$ how can we assess whether it is a {\it bona fide} Wigner function?; (ii) how can we tell whether $F^W (x,p)$ represents a pure or a mixed state?; and finally, (iii) can we reconstruct the wavefunction or the density matrix from the Wigner function? Partial answers to some of these questions have been known for some time. We will review the known results and state what is novel in due course. 

The starting point of this paper is the presentation of several {\it quantum conditions} determining whether a given phase space functional represents a pure or mixed quantum state. These conditions provide a complete classification of phase space functionals in terms of: (i) non-quantum states, (ii) pure quantum states and (iii) mixed quantum states.
One of the pure state quantum conditions presented was originally introduced by Tatarskii \cite{Tatarskii}. For completion we shall review his derivation. 
The rest of the paper is then dedicated to: (1) proving the compatibility between the quantum conditions and the main structures of Wigner quantum mechanics and to assess the degree up to which these structures are in fact determined by the quantum conditions. Namely, we shall prove that: (1a) Moyal (but not classical Liouville) dynamics preserves the pure and mixed quantum conditions, (1b) the solution of a generic stargenvalue equation is a pure state and (1c) the mixed state condition implies the Heisenberg uncertainty relations for general non-commuting observables. (2) Rederiving standard operator quantum mechanics from the phase space formulation. This amounts to: (2a) deriving explicit expressions for wave functions (or density matrixes) associated to a generic pure or mixed state Wigner function, (2b) proving that these wave functions satisfy the proper eigenvalue (Schr\"{o}dinger) equation if the original Wigner function satisfies the corresponding stargenvalue (Moyal) equation. 
The existence of such associated wave functions was established \cite{Baker,Fairlie1} for the energy stargenvalue equation. Here we fully generalize this result by considering the case of arbitrary stargenvalue equations (associated to a generic phase space symbol) and the time dependent case. Moreover, and most importantly, we provide the complete specification of the associated wave functions and density matrix. Finally, (3) we extend the previous results to other phase space quasi-distributions stemming from alternative ordering prescriptions.   

This set of results provides the missing mathematical structure concerning the relation between standard operator and Wigner quantum mechanics and firmly establishes the equivalence of the two formulations.
It also yields a (still missing) complete characterization of the kinematical structure of phase space quantum mechanics.  
This structure is not compatible with classical statistical mechanics. In particular the Liouville time evolution and the classical measurement collapse do not preserve the pure state quantum condition. On the other hand the Moyal theory (dynamics and stargenvalue equations) is fully compatible with the quantum conditions. 
Whether or not the quantum dynamics and the quantum measurement collapse are uniquely determined by the quantum kinematical conditions is an open question to be addressed in a future work \cite{nuno4}. An affirmative answer to this question will imply that quantum mechanics is the only logical statistical formulation for the constrained set of phase space distributions that satisfy the pure state quantum condition.  

This paper is organized as follows. In section 2 we briefly revise some notions concerning pure and mixed states in ordinary quantum mechanics, which we regard as pertinent for the sequel. In section 3 we present a summary of quantum mechanics in phase space. In section 4, we define pure states in phase space and explore some of their properties. In section 5, we derive and study the quantum condition for mixed states. In section 6, we briefly generalize the results to other quasi-distributions. Finally, in sections 7 and 8 we consider two examples: the time independent simple harmonic oscillator and the free Gaussian wave packet. As a by-product we obtain a new (to the best of our knowledge) integral representation for the Hermite polynomials.

Finally, let us make the following remark: for simplicity we shall only consider the case of one-dimensional systems displaying a classical phase space with the structure of a flat manifold. The generalization to higher dimensional systems is straightforward but the extension to curved phase space dynamical systems might only be possible under some restrictive assumptions and, even in that case, may constitute a rather non-trivial problem \cite{nuno4}.

\section{Quantum states}

Let us then consider a one-dimensional system with fundamental observables $\hat q, \hat p$ satisfying Heisenberg's algebra:
\begin{equation}
\left[ \hat q, \hat p \right] = i \hbar.
\end{equation}
The state of the system is completely specified by a vector $| \psi>$ in the Hilbert space or by the so-called density matrix $\hat{\rho} = | \psi>< \psi|$. Let $\left\{ |x>, \hspace{0.2 cm} x \in \Re \right\}$ be the set of eigenstates of $\hat q$. The wavefunction is defined by $\psi (x) = <x| \psi> $. If $\psi (x)$ is square integrable and normalized (with respect to the $L^2(\Re )$ norm) then it is an admissible pure quantum state. We interpret $| \psi (x)|^2 dx$ as the probability of finding the particle in the interval $\left. \right] x, x + dx \left[ \right.$. Likewise, $| \tilde{\psi} (p)|^2 dp$ is the probability that the particle's momentum belong to the interval $\left. \right] p, p + dp \left[ \right.$. Here $\tilde{\psi} (p)$ is the Fourier transform of $\psi (x)$. Planck's constant makes its appearance in eq.(1). The previous considerations then lead to one of the main results of quantum mechanics, namely Heisenberg's uncertainty relations:
\begin{equation}
\Delta x \cdot \Delta p \ge \frac{\hbar}{2},
\end{equation}
where $\left( \Delta z \right)^2 \equiv < z^2> - < z>^2$.

Now, let $\hat A$ be some hermitian operator functional of $\hat q$ and $\hat p$ with discrete nondegenerate spectrum\footnote{The generalization to continuous and degenerate spectra is straightforward.}. Suppose further that $ \left\{|a>, \hspace{0.2 cm} a \in I \right\}$ is a complete orthonormal set of eigenvectors of $\hat A$:
\begin{equation}
\hat A |a> = a |a>, \hspace{0.5 cm} a \in I.
\end{equation}
Then the wavefunction can be expanded as follows:
\begin{equation}
| \psi > = \sum_{a \in I} C_a |a>,
\end{equation}
where $\left\{C_a, \hspace{0.2 cm} a \in I \right\}$ are a set of complex constants, such that $\sum_{a \in I} |C_a|^2 =1$. The probability that a measurement of the observable $\hat A$ yield the eigenvalue $a$ is $|C_a|^2$.

Alternatively, suppose that instead of a single state $| \psi>$, we have a statistical mixture of states $\left\{| \psi^{(\beta)}>, \hspace{0.2 cm} \beta \in B \right\}$, each one with probability $p_{\beta}$. In this case the density matrix becomes:
\begin{equation}
\hat \rho = \sum_{\beta \in B} p_{\beta} | \psi^{(\beta)}><\psi^{(\beta)}| = \sum_{\beta \in B} \sum_{a,b \in I} p_{\beta} C_a^{(\beta)} C_b ^{(\beta) *} |a><b| = \sum_{a,b \in I} \rho_{ab} |a><b|,
\end{equation}
where 
\begin{equation}
0 \le p_{\beta} \le 1, \hspace{0.5 cm} \forall \beta \in B ; \hspace{1 cm} \sum_{\beta \in B} p_{\beta} =1,
\end{equation}
and 
\begin{equation}
\left\{
\begin{array}{l}
| \psi^{(\beta)}> = \sum_{a \in I} C_a ^{(\beta)} |a>, \hspace{1 cm} \sum_{a \in I} |C_a^{(\beta)}|^2 =1, \hspace{0.5 cm} \forall \beta \in B \\
\\
\rho_{ab} \equiv \sum_{\beta \in B} p_{\beta} C_a^{(\beta)} C_b^{(\beta) *}, \hspace{0.5 cm} a,b \in I.
\end{array}
\right.
\end{equation}
From these relations we can easily infer the following properties of $\hat{\rho}$: 

\vspace{0.3 cm}

1) $\hat{\rho}$ is hermitian;

2) $\hat{\rho}$ is non negative:
\begin{equation}
< \phi | \hat{\rho} | \phi> \hspace{0.2 cm} \ge 0,
\end{equation}
for any state $| \phi>$ in the Hilbert space; and

3) $\hat{\rho}$ is normalized:
\begin{equation}
Tr \hat{\rho} =1.
\end{equation}
Conversely, it can be shown \cite{Messiah,Reed,Landsman} that if a matrix satisfies properties 1) - 3), then it admits the following decomposition:
\begin{equation}
\hat{\rho} = \sum_{\beta \in B} p_{\beta} | \psi^{(\beta)}>< \psi^{(\beta)} |,
\end{equation}
where $\left\{ | \psi^{(\beta)}>, \beta \in B \right\}$ is some set of normalized states and $p_{\beta}$ are some real constants satisfying (6). Equation (10) means that the density matrix is inevitably a superposition of pure states. 
Conditions 1) - 3) allow for a probabilistic interpretation of the diagonal entries of the density matrix, which is independent of the complete orthonormal basis of the Hilbert space.

It should also be noticed that we can always find a decomposition (10) where the states $|\psi^{(\beta)}>$ are orthogonal, i.e. $<\psi^{(\alpha)}|\psi^{(\beta)}>=\delta_{\alpha,\beta}$. In this case the density matrix is diagonal and the states $|\psi^{(\beta)}> $ are eigenstates of $\hat{\rho}$. The associated eigenvalues $p_{\beta}$ are non-negative and normalized by virtue of (8) and (9). Using this particular decomposition one can easily prove that:

i) A hermitian and normalized operator $\hat{\rho}$ (i.e. satisfying properties 1) and 3)) is a pure state density matrix iff it satisfies $\hat{\rho}=\hat{\rho}\hat{\rho}$. In fact property 2) follows immediately from the preceding properties. Hence, $\hat{\rho}$ satisfies 1) to 3) and can be written in the form (10) where $p_{\beta}$ satisfies (6) and $<\psi^{(\alpha)}|\psi^{(\beta)}>=\delta_{\alpha,\beta}$. We then have:
\begin{eqnarray}
\hat{\rho}=\hat{\rho}\hat{\rho} & \Longleftrightarrow & \sum_{\beta \in B} p_{\beta} |\psi^{(\beta)}><\psi^{(\beta)}|=
\sum_{\beta \in B} p_{\beta}^2 |\psi^{(\beta)}><\psi^{(\beta)}| \Longleftrightarrow p_{\beta}^2=p_{\beta} , \, \forall \beta \in B \nonumber \\
& \Longleftrightarrow & p_{\beta}=\delta_{\beta,\beta_0} \, \mbox{ for some $\beta_0 \in B$},
\end{eqnarray}
where we used eq.(6). Hence, $\hat{\rho}=|\psi^{(\beta_0)}><\psi^{(\beta_0)}|$ and so it is a pure state.

We leave as a simple exercise to further prove that:
 
ii) If $\hat{\rho}$ is a density matrix (i.e. it satisfies properties 1) to 3)) then $\hat{\rho}$ is a pure state 
iff $Tr \hat{\rho}\hat{\rho}=1$. 

iii) If $\hat{\rho}$ is a density matrix then it is a strictly mixed state iff $Tr \hat{\rho}\hat{\rho} < 1$.

This completes our review of some well-known results of the density matrix formulation of quantum mechanics. This section can then be summarized as follows. If $\hat{\rho}$ is a matrix which satisfies properties 1) - 3) then it is an admissible quantum state. If it further satisfies ii) (respectively iii)) then is a pure (strictly mixed) quantum state. Alternatively, if it satisfies 1),3) and i) then it is a pure quantum state.

\section{Weyl-Wigner formulation of quantum mechanics}

In this section we revise the Weyl-Wigner formulation of quantum mechanics. Let us start with Weyl's correspondence rule. The Weyl map $V_W$ is an isomorphism from the quantum algebra $\hat{{\cal A}}$ of linear operators with an operator product $\cdot$ and a commutator $\left[ , \right]$ to the classical algebra ${\cal A}$ of phase space functions with a non-commutative $*$-product (the Groenewold product \cite{Groenewold}) and a bracket (the Moyal bracket \cite{Moyal}). It is defined by: 
\begin{equation}
\hat A \left( \hat q, \hat p \right) \mapsto V_W (\hat A) = A(x,p) = \frac{\hbar}{2 \pi} \int d \xi \int d \eta \hspace{0.2 cm} Tr \left\{ \hat A \left( \hat q, \hat p \right) e^{i \xi \hat q + i \eta  \hat p} \right\} e^{-i \xi x - i \eta p}.
\end{equation} 
The definitions of the non-commutative $*$-product and the Moyal bracket follow immediately:
\begin{equation}
\left\{
\begin{array}{l l}
A * B & \equiv  V_W \left( \hat A \cdot \hat B \right)\\
\left[A,B \right]_M & \equiv  \frac{1}{i \hbar} \left(A *B - B *A \right) = \frac{1}{i \hbar} V_W \left( \left[ \hat A, \hat B \right] \right)
\end{array} 
\right.
\end{equation}
From equation (12), we get the following formulae:
\begin{equation}
\left\{
\begin{array}{l l}
A * B & = A (x, p) e^{\frac{i \hbar}{2} {\hat{\cal J}}}  B (x, p) =A(x,p-i\frac{\hbar}{2} \frac{ {\buildrel { \rightarrow}\over\partial}}{\partial x}) B(x,p+i\frac{\hbar}{2} \frac{ {\buildrel { \leftarrow}\over\partial}}{\partial x})\\
& \\
& = \frac{1}{\hbar^2 \pi^2} \int dp'dp''dx'dx''A(x',p') B(x'',p'') e^{\frac{-2i}{\hbar } \{ p(x'-x'') + p'(x''-x)+p''(x-x') \} }\\
& \\
\left[ A , B \right]_M & = \frac{2}{\hbar}  A (x, p) \sin \left(\frac{\hbar}{2} {\hat{\cal J}} \right) B ( x, p),
\end{array}
\right.
\end{equation}
where ${\hat{\cal J}}$ is the "{\it Poisson}" operator: 
\begin{equation}
{\hat{\cal J}} \equiv   \frac{ {\buildrel { \leftarrow}\over\partial}}{\partial x} \frac{ {\buildrel { \rightarrow}\over\partial}}{\partial p} -  \frac{{\buildrel {\leftarrow}\over\partial}}{\partial p}  \frac{{\buildrel { \rightarrow}\over\partial}}{\partial x},
\end{equation}
the derivatives ${\buildrel { \leftarrow}\over\partial}$ and ${\buildrel { \rightarrow}\over\partial}$ acting on $A$ and $B$, respectively. 
The Weyl map of the density matrix is proportional to the Wigner function \cite{Wigner}:
\begin{equation}
\begin{array}{c}
F^W(x,p,t) \equiv \frac{1}{2 \pi \hbar} V_W \left( \hat{\rho} (t) \right) = \frac{1}{(2 \pi)^2} \int d \xi \int d \eta \hspace{0.2 cm} Tr \left\{ \hat{\rho} (t) e^{i \xi \hat q + i \eta \hat p} \right\} e^{-i \xi x - i \eta p} = \\
\\
= \frac{1}{2 \pi} \int dy \hspace{0.2 cm} e^{-i y p} < x+ \frac{\hbar y}{2}| \hat{\rho} (t) | x - \frac{\hbar y}{2}>.
\end{array}
\end{equation}
In particular for a pure state, $\hat{\rho} = | \psi>< \psi|$, we get:
\begin{equation}
F^W_{pure}(x,p,t) = \frac{1}{2 \pi} \int dy \hspace{0.2 cm} e^{- i y p} \psi^* \left( x - \frac{\hbar y}{2} , t \right) \psi \left( x + \frac{\hbar y}{2} , t \right).
\end{equation} 
The pure state Wigner function is square integrable, normalized (in the sense that $\int dxdy F^W_{pure}(x,p,t)=1$) and real, but, in general, it is not positive defined. The mean value of the operator $\hat A$ can be computed according to the following appealing formula:
\begin{equation}
< \hat A (t)> = \int dx \int dp \hspace{0.2 cm} A (x,p) F^W (x,p,t).
\end{equation}
If the operator $\hat A$ has nondegenerate spectrum and $|a>$ is one of its eigenvectors 
(3), then the corresponding Wigner function $F_a^W (x,p) = \frac{1}{2 \pi \hbar} V_W \left( |a><a| \right)$ is the solution of the following $*$-genvalue equations \cite{Fairlie1,Dias2}:
\begin{equation}
A(x,p) * F_a^W (x,p) = a F_a^W (x,p), \hspace{1 cm} F_a^W (x,p) * A (x,p) = a F_a^W (x,p).
\end{equation} 
The probability that a measurement of the observable $\hat A$ yield the eigenvalue $a$ is then given by
\begin{equation}
{\cal P} \left( A (t) = a \right) = \int dx \int dp \hspace{0.2 cm} F^W (x,p ,t) F_a^W (x,p),
\end{equation}
provided $F^W$ is normalized. If the spectrum is degenerate, then the generalization is straightforward, \cite{Dias2}. 

If the state is mixed, then from (5), we have:
\begin{equation}
F^W (x,p,t) = \sum_{\beta \in B} p_{\beta} F^W_{pure,\beta}(x,p,t)=\sum_{a,b \in I} \rho_{ab} (t)  F_{ab}^W  (x,p),
\end{equation}
where the non-diagonal Wigner function $F_{ab}^W$ reads:
\begin{equation}
F_{ab}^W (x,p) = \frac{1}{2 \pi} \int dy \hspace{0.2 cm} e^{-iyp} \psi_b^* \left( x - \frac{\hbar  y}{2} \right) \psi_a \left( x + \frac{\hbar y}{2} \right),
\end{equation}
the pure state Wigner functions $F^W_{pure,\beta}(x,p,t)$ are of the form (17) (and can always be chosen so that they satisfy $F^W_{pure,\beta}(x,p,t)*F^W_{pure,\alpha}(x,p,t)=\frac{\delta_{\beta,\alpha}}{2\pi \hbar} F^W_{pure,\beta}(x,p,t)$), $p_{\beta}$ satisfies (6) and $\rho_{ab}$ is given by (7) thus satisfying the following conditions: i) $\rho_{ab}=\rho_{ba}^*$, ii) $\rho_{aa} \ge 0 , \forall a \in I$ and iii) $\sum_{a \in I} \rho_{aa} =1$.
 
If the wavefunctions $\psi_1$, $\psi_2$ are solutions of the Schr\"odinger equation,
\begin{equation}
i \hbar \frac{\partial \psi_j}{\partial t} = - \frac{\hbar^2}{2m} \frac{\partial^2 \psi_j}{\partial x^2} + V (x) \psi _j, \hspace{0.5 cm} j=1,2,
\end{equation}
then the non-diagonal Wigner function $F_{12}^W$ obeys the Moyal equation:
\begin{equation}
\frac{\partial F_{12}^W}{\partial t} = \left[ H , F_{12}^W \right]_M,
\end{equation}
with 
\begin{equation}
H (x,p) = \frac{p^2}{2m} +  V(x).
\end{equation}  
Likewise if $\psi_j$ are energy eigenstates with eigenvalues $E_j$ $(j=1,2)$,
\begin{equation}
- \frac{\hbar^2}{2m} \frac{\partial^2 \psi_j}{\partial x^2} + V(x) \psi_j = E_j \psi_j, \hspace{0.5 cm} j=1,2,
\end{equation}
then $F_{12}^W$ is a solution of the $*$-genvalue equations:
\begin{equation}
H(x,p) * F_{12}^W = E_1 F_{12}^W , \hspace{1 cm} F_{12}^W * H(x,p) = E_2 F_{12}^W.
\end{equation} 
This completes our overview of the Weyl-Wigner formulation of quantum mechanics.

\section{Pure states in quantum phase space}

In this section we will start to address the equivalence of the standard operator and Wigner's formulations of quantum mechanics. We shall first concentrate on pure states. Let us start by presenting the relevant definitions. A phase space function is a {\it pure quantum state} iff is of the form (17) for some complex valued, square integrable and normalized function $\psi(x,t)$. In other words, iff is the Weyl transform of a pure state density matrix. A phase space function of the form (22) for some complex valued, normalized functions $\psi_a(x) \not= \psi_b(x)$ will be designated by {\it non-diagonal pure state}. They will be important for the study of mixed states in the next section. When there is no risk of confusion we will also designate them by just pure states. Finally, we will also consider phase space functionals of the form (22) where the associated wave functions are distributions. These states are non-physical (even if real the associated wave functions do not belong to the Hilbert space) and will be important (as in standard operator quantum mechanics) because they provide basis for the space of physical phase space functionals. We will designate them by {\it distributional (non-diagonal, if complex) pure states}. 

Let then $F^W (x,p,t)$ be some function in quantum phase space. How do we know whether this function represents a pure quantum state? i.e. how do we know that this function is of the form (17)? The two following Lemmas provide two alternative characterizations of phase space pure states.

\vspace{0.3 cm}
\noindent
{\underline{\bf Lemma 1:}} A real and normalized phase space function is a pure state Wigner function iff it satisfies:
\begin{equation}
F^W_{pure}*F^W_{pure}=\frac{1}{2\pi \hbar} F^W_{pure}.
\end{equation}

\vspace{0.3 cm}
\noindent
{\underline{\bf Proof:}} $F^W_{pure}$ is a pure state Wigner function iff it is the Weyl transform of a pure state density matrix $|\psi><\psi|$. By applying the Weyl map to the conditions 1), 3) and i) of section 2 we immediately get that $F^W_{pure}$ is: 1) real, 2) normalized and 3) satisfies (28). 

Conversely, if a phase space function $F$ satisfies (28) then by virtue of Theorem 2 (see ahead) $F$ is of the form (22). If in addition $F$ is real and normalized then $F$ is of the form (17) for some normalized wave function $\psi$. Hence, $F$ is a pure state.$_{\Box}$

Equation (28) provides an elegant, compact form for the pure state quantum condition. However, the evaluation of the starproduct in (28) may constitute in practice a rather difficult problem. Following Tatarskii \cite{Tatarskii} we now provide an alternative, more practical pure state quantum condition:    

\vspace{0.3 cm}
\noindent
{\underline{\bf Lemma 2:}} Let $F^W(x,p,t)$ be square integrable and let us define the following function:
\begin{equation}
Z (x,j,t) = \int dp \hspace{0.2 cm} e^{ijp} F^W (x,p,t).
\end{equation}
If $Z (x,j,t)$ satisfies the nonlinear partial differential equation,
\begin{equation}
\frac{\partial^2}{\partial j^2} \ln Z (x,j,t) = \left( \frac{\hbar}{2} \right)^2 \frac{\partial^2}{\partial x^2} \ln Z (x,j,t),
\end{equation}
then $F^W (x,p,t)$ is phase space function of the form:
\begin{equation}
F_{12}^W (x,p,t) = \frac{1}{2 \pi} \int dy \hspace{0.2 cm} e^{-iyp} \psi_2^* \left( x - \frac{\hbar  y}{2},t \right) \psi_1 \left( x + \frac{\hbar y}{2} ,t\right),
\end{equation} 
where $\psi_1$, $\psi_2$ are some complex square integrable functions. If in addition $F^W_{12}$ is real and normalized then it is a pure state Wigner function. Conversely, if $F^W(x,p,t)$ is a pure state (possibly non-diagonal) then it satisfies eq.(30).

\vspace{0.3 cm}
\noindent
{\underline{\bf Proof:}} Eq.(30) can be seen as a wave equation, where the "velocity" is $c= \frac{\hbar}{2}$, if $j$ plays the role of "time". Its solution is:
\begin{equation}
\ln Z (x,j,t) = \ln \psi_2^* \left( x -  \frac{\hbar  j}{2} ,t\right) + \ln \psi_1 \left( x + \frac{\hbar j}{2},t \right),
\end{equation}
where $\psi_1$, $\psi_2$ are arbitrary complex functions. On the other hand, $Z (x,j,t)$ is the Fourier transform of $F^W (x,p,t)$. And so we get:
\begin{equation}
F^W (x,p,t) = \frac{1}{2 \pi} \int dy \hspace{0.2 cm} e^{-iyp} Z (x,y,t) = \frac{1}{2 \pi} \int dy \hspace{0.2 cm} e^{-iyp} \psi_2^* \left( x - \frac{\hbar  y}{2} ,t\right) \psi_1 \left( x + \frac{\hbar y}{2} ,t\right),
\end{equation} 
and so $F^W$ is indeed of the form (31). Moreover and since $F^W$ is square integrable so are $\psi_1$ and $\psi_2$. If we further impose that besides being a solution of (30), $F^W (x,p,t)$ should also be real and normalized, then it is trivial to verify that $F^W (x,p,t)$ is indeed of the form (17), where $\psi (x)$ is normalized. Finally, the converse result follows immediately from the definition of $Z(x,j,t)$.$_{\Box}$

We shall call eq.(30) the {\it pure state quantum condition}. Its real and normalized solutions are pure state Wigner functions. A solution of eq.(30) is a physical state iff is a pure state. The non-real solutions, on the other hand, are the non-diagonal pure states. Equation (30) was originally introduced by Tatarskii \cite{Tatarskii}. Let us then explore some properties of the quantum pure states.\\
\\ 
{\underline{\bf Theorem 1:}} Let $F^W(x,p,t)$ satisfy the quantum condition (30) at the initial time $t=0$ and let its time evolution be given by the Moyal equation (24). Then $F^W(x,p,t)$ satisfies (30) for all times.\\
\\
{\underline{\bf Proof:}} This theorem was also originally stated by Tatarskii \cite{Tatarskii}. Let us briefly review his proof: if $F^W(x,p,0)$ satisfies eq.(30) then it is given by eq.(31) for some complex functions $\psi_1(x,0)$ and $\psi_2(x,0)$. It is well known \cite{Moyal} that if $\psi_1(x,t)$, $\psi_2(x,t)$ are the Schr\"odinger time evolution of $\psi_1(x,0)$ and $\psi_2(x,0)$ then the associated Wigner function $F^W(x,p,t)$ (given by eq.(31)) satisfies the Moyal equation for the Weyl symbol of the original Hamiltonian operator. Since the Moyal time evolution is uniquely determined by the initial time Wigner function $F^W(x,p,0)$, we conclude that $F^W(x,p,t)$ is of the form (31) for all times.$_{\Box}$

Later in this section we will show that this property is not shared by the classical (Liouville) time evolution. The next theorem extends Baker's converse construction to the solution of a generic stargenvalue equation.\\
\\
{\underline{\bf Theorem 2:}} Let $A(x,p)=V_W(\hat A)$ be a generic phase space symbol associated to a generic linear operator $\hat A$. Then the solution of the stargenvalue equations:
\begin{equation}
A(x,p) * F^W(x,p)=a F^W(x,p), \qquad F^W(x,p) * A(x,p)= b F^W(x,p),
\end{equation}
for $a,b$ in the spectrum of $\hat A$, is a ({\it distributional}, if the spectrum of $\hat A$ is continuous) pure state of
the form (31). The associated wave functions $\psi_1(x), \psi_2(x)$ satisfy the eigenvalue equations $\hat A \psi_1(x) = a \psi_1(x)$ and $\hat A^{\dagger} \psi_2(x) = b^* \psi_2(x)$.
 
\vspace{0.3 cm}
\noindent
{\underline{\bf Proof:}} A generic linear operator can be displayed in a fully symmetrized form through the Weyl prescription:
\begin{equation}
\hat A(\hat q,\hat p)=  \int d \xi \int d \eta \hspace{0.2 cm} \alpha(\xi ,\eta ) e^{i \xi \hat q + i \eta \hat p},
\end{equation} 
where $\alpha (\xi,\eta)=\frac{\hbar}{2 \pi} Tr \left\{ \hat A \left( \hat q, \hat p \right) e^{-i \xi \hat q - i \eta  \hat p} \right\}$. The Weyl transform of $\hat A$ is given by (cf.(12)):
\begin{equation}
A(x,p)= \int d \xi \int d \eta \hspace{0.2 cm} \alpha(\xi,\eta) e^{i \xi x + i \eta p}.
\end{equation} 
We now assume that $F^W(x,p)$ satisfies the stargenvalue equation (34) with $A(x,p)$ given by (36). Using eqs.(14,29) we get for the first stargenvale equation (34):
\begin{eqnarray}
a F^W(x,p) & = & \frac{1}{2\pi} \int d \xi \int d \eta \int dy \hspace{0.2 cm} \alpha(\xi,\eta) e^{i \xi x + i \eta (p-i\frac{\hbar}{2} {\buildrel { \rightarrow}\over\partial}_x)} e^{-iy(p+i\frac{\hbar}{2} {\buildrel { \leftarrow}\over\partial}_x )} Z(x,y) \nonumber \\
&=& \frac{1}{2\pi} \int d \xi \int d \eta \int dy \hspace{0.2 cm} \alpha(\xi,\eta) e^{i \xi (x +\hbar y/2)} \left[e^{-\eta {\buildrel { \rightarrow}\over\partial}_y} e^{-iyp} \right] e^{\eta \frac{\hbar}{2} {\buildrel { \rightarrow}\over\partial}_x} Z(x,y) \nonumber \\
&=& \frac{1}{2\pi} \int d \xi \int d \eta \int dy \hspace{0.2 cm} \alpha(\xi,\eta) e^{i \xi x -iyp} e^{\eta {\buildrel { \rightarrow}\over\partial}_y} e^{i\hbar \xi y /2} e^{-\eta {\buildrel { \rightarrow}\over\partial}_y} e^{\eta {\buildrel { \rightarrow}\over\partial}_y}
e^{\eta \frac{\hbar}{2} {\buildrel { \rightarrow}\over\partial}_x} Z(x,y), 
\end{eqnarray}
where in the last step we integrated by parts in $y$ and used the relation $e^{-\eta {\buildrel { \rightarrow}\over\partial}_y}e^{\eta {\buildrel { \rightarrow}\over\partial}_y}=1$. The Baker-Campbell-Hausdorff relation yields:
\begin{equation}
e^{\eta {\buildrel { \rightarrow}\over\partial}_y} e^{i\hbar \xi y /2} e^{-\eta {\buildrel { \rightarrow}\over\partial}_y}=
e^{i \hbar \xi \eta /2} e^{i \hbar \xi y /2} ,
\end{equation}
and substituting in (37) we get:
\begin{equation}
a F^W(x,p)=\frac{1}{2\pi} \int d \xi \int d \eta \int dy \hspace{0.2 cm} \alpha(\xi,\eta) e^{i \xi (x +\hbar y/2)} e^{-iyp} e^{i \hbar \xi \eta /2} 
e^{\eta \hbar (\frac{1}{2}{\buildrel { \rightarrow}\over\partial}_x +\frac{1}{\hbar } {\buildrel { \rightarrow}\over\partial}_y) } Z(x,y) .
\end{equation}
We notice that the commutator of $\hat C=i \xi (x +\hbar y/2)$ and $\hat D=\eta \hbar (\frac{1}{2}{\buildrel { \rightarrow}\over\partial}_x +\frac{1}{\hbar } {\buildrel { \rightarrow}\over\partial}_y)$ is a $c$-number and thus $e^{\hat C} e^{\hat D}=e^{\hat C+\hat D} e^{\frac{1}{2} [\hat C,\hat D]}$. Substituting in (39) we get:
\begin{equation}
a F^W(x,p)= \frac{1}{2\pi} \int d \xi \int d \eta \int dy \hspace{0.2 cm} \alpha(\xi,\eta) e^{-iyp} e^{i \xi (x +\hbar y/2)   
+ i \eta \left[ -i \hbar (\frac{1}{2}{\buildrel { \rightarrow}\over\partial}_x +\frac{1}{\hbar } {\buildrel { \rightarrow}\over\partial}_y ) \right]} Z(x,y) .
\end{equation}
Using (36) and (29) we finally obtain:
\begin{equation}
\hat A\left[ x +\frac{\hbar}{2} y, -i \hbar (\frac{1}{2}{\buildrel { \rightarrow}\over\partial}_x +\frac{1}{\hbar } {\buildrel { \rightarrow}\over\partial}_y ) \right] Z(x,y)=a Z(x,y).
\end{equation}
Following exactly the same steps we get for the second equation in (34):
\begin{equation}
\hat A\left[ x -\frac{\hbar}{2} y, i \hbar (\frac{1}{2}{\buildrel { \rightarrow}\over\partial}_x -\frac{1}{\hbar } {\buildrel { \rightarrow}\over\partial}_y ) \right] Z(x,y)=b Z(x,y).
\end{equation}
Let us now define the new variables:
\begin{equation}
\left\{ \begin{array}{lll}
u &=& x+ \frac{\hbar}{2} y \\
v &=& x-\frac{\hbar}{2} y 
\end{array} \right. \quad \Longrightarrow \quad
\left\{ \begin{array}{lll}
{\buildrel { \rightarrow}\over\partial}_u &=& \frac{1}{2} {\buildrel { \rightarrow}\over\partial}_x + \frac{1}{\hbar} {\buildrel { \rightarrow}\over\partial}_y  \\
{\buildrel { \rightarrow}\over\partial}_v &=& \frac{1}{2} {\buildrel { \rightarrow}\over\partial}_x - \frac{1}{\hbar}
{\buildrel { \rightarrow}\over\partial}_y 
\end{array} \right. ,
\end{equation}
and the new function $G(u,v)=Z(\frac{u+v}{2},\frac{u-v}{2})=Z(x,y)$. From eqs.(41,42) we have:
\begin{equation}
\left\{ \begin{array}{lll}
\hat A\left[ u, -i \hbar {\buildrel { \rightarrow}\over\partial}_u \right] G(u,v) & = & a G(u,v) \\
\hat A\left[ v, i \hbar {\buildrel { \rightarrow}\over\partial}_v \right] G(u,v) & = & b G(u,v). 
\end{array} \right.
\end{equation}
From where it follows that $G(u,v)=\phi(v) \psi(u)$ where $\psi(u)$ satisfies:
\begin{equation}
\hat A\left[ u, -i \hbar {\buildrel { \rightarrow}\over\partial}_u \right] \psi(u)  =  a \psi(u), 
\end{equation}
and $\phi(v)$ obeys to:
\begin{equation}
\hat A\left[ v, i \hbar {\buildrel { \rightarrow}\over\partial}_v \right] \phi(v)  =  b \phi(v) \Longleftrightarrow
\hat A^{\dagger}\left[ v,  -i\hbar {\buildrel { \rightarrow}\over\partial}_v \right] \phi^*(v)  =  b^* \phi^*(v) .
\end{equation}
Let $\psi_2(x)=\phi^*(x)$ and $\psi_1(x)=\psi(x)$. Then $Z(x,y)=G(u,v)=\psi_2^*(v)\psi_1(u)=\psi_2^*(x-\frac{\hbar}{2}y)\psi_1(x+\frac{\hbar}{2}y)$ and thus $F^W(x,p)$ is a (possibly distributional) pure state. Furthermore, $\psi_1$ and $\psi_2$ satisfy the proper eigenvalue equations. Finally if $\hat A$ is hermitian then eq.(46) reduces to $ \hat A\left[ x,  -i\hbar {\buildrel { \rightarrow}\over\partial}_x \right] \psi_2(x)  =  b \psi_2(x)$.$_{\Box}$

Later in this section we will provide the specification of the wave functions $\psi_1,\psi_2$ in terms of the original Wigner function. First let us explore the physical meaning of the "initial conditions" for eq.(30). We define the mean $n$-th order momentum as:
\begin{equation}
< p^n (x,t) >_{12} \equiv \frac{\int dp \hspace{0.2 cm} p^n F_{12}^W (x,p,t)}{{\cal P}_{12} (x,t)},
\end{equation}
where
\begin{equation}
{\cal P}_{12} (x,t) = \int dp \hspace{0.2 cm} F_{12}^W (x,p,t) = \psi_2^* (x,t) \psi_1 (x,t).
\end{equation}
The function $Z (z,j,t)$ then works as a generating function of the higher momentum correlations $\Delta_{12}^{(n)} p (x,t)$:
\begin{equation}
\left\{
\begin{array}{l}
{\cal P}_{12} (x,t) = Z (x,0,t),\\
\left. \Delta_{12}^{(n)} p (x,t) = \frac{1}{i^n} \frac{\partial^n}{\partial j^n} \ln Z (x,j,t) \right|_{j=0}.
\end{array}
\right.
\end{equation}
In particular, we have:
\begin{equation}
\left\{
\begin{array}{l}
\Delta_{12}^{(1)} p (x,t) \equiv < p(x,t)>_{12},\\
\Delta_{12}^{(2)} p (x,t) \equiv < p^2 (x,t) >_{12} - < p (x,t)>_{12}^2.
\end{array}
\right.
\end{equation}
From eq.(30,49) we can obtain the following recursive relation:
\begin{equation}
\Delta_{12}^{(n+2)} p (x,t) = \left( \frac{\hbar}{2i} \right)^2 \frac{\partial^2}{\partial x^2} \Delta_{12}^{(n)} p (x,t), \hspace{0.5 cm} \forall n \ge 1.
\end{equation}
This means that we can compute all $\Delta_{12}^{(n)} p$ from $\Delta_{12}^{(1)} p$ and $\Delta_{12}^{(2)} p$. Consequently, we get for $n \ge 1$:
\begin{equation}
\left\{
\begin{array}{l l}
\Delta_{12}^{(n)} p (x,t) = \left( \frac{\hbar}{2i} \right)^n \frac{\partial^n}{\partial x^n} \ln {\cal P}_{12} (x,t), & \mbox{if $n$ is even}\\
 & \\
\Delta_{12}^{(n)} p (x,t) = \left( \frac{\hbar}{2i} \right)^{n-1} \frac{\partial^{n-1}}{\partial x^{n-1}} <p (x,t) >_{12} , & \mbox{if $n$ is odd}
\end{array}
\right.
\end{equation}
where 
\begin{equation}
< p(x,t)>_{12} = \frac{\hbar i}{2} \frac{\partial}{\partial x} \ln \left[ \frac{ \psi_2^* (x,t)}{\psi_1 (x,t)} \right].
\end{equation}
A pure state could be seen as one where all the higher momentum correlations depend only upon $\Delta^{(1)} p $ and $\Delta^{(2)} p $. This is not surprising since ${\cal P}_{12}(x,t)$ and $< p(x,t)>_{12}$ provide suitable "initial conditions" for eq.(30) and thus completely determine the resulting pure state. 

We proved the consistency of the pure state quantum condition with both the Moyal and stargenvalue equations and studied the physical meaning of the "initial conditions of eq.(30). We now address the issue of deriving explicit expressions for the wave functions in terms of the original pure state Wigner function. This will be the purpose of the two following subsections.
      
\subsection{Time independent pure states}

Let us assume that the time independent function $F^W_{12} (x,p)$ is square integrable and satisfies eq.(30). Then we already know that $F_{12}^W$ takes the form (31). How can we construct the functions $\psi_1$, $\psi_2$ from the knowledge of $F_{12}^W$? The answer is trivial:
\begin{equation}
\left\{
\begin{array}{l}
\psi_1 (x) = N_1 \int dp \hspace{0.2 cm} e^{i p x / \hbar} F_{12}^W \left( \frac{x}{2} , p \right) = N_1 Z \left( \frac{x}{2}, \frac{x}{\hbar} \right),\\
\\  
\psi_2 (x) = N_2 \int dp \hspace{0.2 cm} e^{i p x / \hbar} \left[ F_{12}^W \left( \frac{x}{2} , p \right) \right]^* = N_2 Z^* \left( \frac{x}{2}, \frac{x}{\hbar} \right),
\end{array}
\right.
\end{equation}
where $N_1$, $N_2$ are some complex constants (obviously $N_1 = [\psi_2^* (0)]^{-1}$ and $N_2 = [\psi_1 (0)]^{-1}$). Furthermore and since $Z \left( \frac{x}{2}, \frac{x}{\hbar} \right)$ is a square integrable function, then so are $\psi_1$, $\psi_2$. There may be a problem if $\psi_1 (0) =0$ or $\psi_2 (0)=0$. However, this can be easily overcome, as we now argue. Suppose that when we compute $\psi_1 (x) $ according to equation (54), we get zero identically. This is a sign that $\psi_2 (0)=0$. In that case, we can use the alternative formula:
\begin{equation}
\psi_1 (x) = M_1 \int dp \hspace{0.2 cm} e^{ipx / \hbar} \frac{\partial }{\partial x} F_{12}^W \left( \frac{x}{2} , p \right),
\end{equation}
where $M_1$ is another complex constant. Indeed, from (31), the right-hand side of (55) reads:
\begin{equation}
\frac{M_1}{2} \left[\psi_2'^* (0) \psi_1 (x) + \psi_2^* (0) \psi_1' (x)  \right].
\end{equation}
In all cases of interest, the function $\psi_2 (x)$ and its derivative $\psi_2'(x)$ do not vanish simultaneously at the same point\footnote{For a solution of the Schr\"odinger equation, this would imply that $\psi_2 (x)$ vanish identically.}. We then have $\psi_2 (0)=0$, but $\psi_2'(0) \ne 0$, and hence (56) yields:
\begin{equation}
\frac{M_1}{2} \psi_2'^* (0) \psi_1 (x).
\end{equation}
And thus the previous arguments still apply. These technicalities could be formally averted, if instead of evaluating $\psi_1$ or $\psi_2$ at $x=0$, we had chosen some other point $x_0$ such that $\psi_1 (x_0) \ne 0 $ and $\psi_2 (x_0) \ne 0$. However it turns out that in most cases, the calculation is easier to perform at $x=0$ (see section 7).

In theorem 2 we proved that if $F^W_{12}$ satisfies the generic $*$-genvalue equations (34) then $F^W_{12}$ is a pure state and the associated wave functions satisfy the corresponding eigenvalue equations. Now that we have explicit formulae for the associated wave functions let us show that they provide the proper eigenfunctions for the case of the energy $*$-genvalue equation:

\vspace{0.3 cm}
\noindent
{\underline{\bf Theorem 3:}} If $F_{12}^W$ obeys the $*$-genvalue equations (27), then the associated wave functions $\psi_1$, $\psi_2$ (eq.(54)) satisfy the corresponding energy eigenvalue equations (eq.(26)). 

\vspace{0.3 cm}
\noindent
{\underline{\bf Proof:}}
From the first equation in (27) and from (14), we get:
\begin{equation}
\left\{\frac{1}{2m} \left( p^2  - i \hbar p \frac{\partial }{\partial x} - \frac{\hbar^2}{4} \frac{\partial^2 }{\partial x^2} \right) + \sum_{n=0}^{+ \infty} \frac{1}{n!} \left( \frac{i \hbar}{2} \right)^n V^{(n)} (x) \frac{\partial^n }{\partial p^n} \right\} F_{12}^W (x,p) = E_1 F_{12}^W (x,p),
\end{equation}
where $V^{(n)} (x)$ is the $n$-th derivative of $V(x)$. If we perform the substitution $x \to \frac{x}{2}$, multiply the equation by $N_1 e^{ipx / \hbar}$ and integrate over $p$, we get: $A_1 + A_2 = A_3$, where:
\begin{equation}
\begin{array}{c}
A_1 (x) = \frac{N_1}{2m} \int d p \hspace{0.2 cm} e^{i p x / \hbar} \left(p^2  - 2 i \hbar p \frac{\partial }{\partial x} - \hbar^2 \frac{\partial^2 }{\partial x^2} \right) F_{12}^W \left(\frac{x}{2}, p \right) = \\
\\
= - \frac{N_1 \hbar^2}{2m} \int d p \hspace{0.2 cm} \frac{\partial^2 }{\partial x^2} \left[e^{i p x / \hbar} F_{12}^W \left(\frac{x}{2}, p \right) \right] = - \frac{\hbar^2}{2m} \frac{d^2 }{d x^2} \psi_1 (x) ,
\end{array}
\end{equation}

\vspace{0.5 cm}

\begin{equation}
\begin{array}{c}
A_2 (x) = N_1  \int d p \hspace{0.2 cm}  e^{i p x / \hbar} \sum_{n=0}^{+ \infty} \frac{1}{n!} \left( \frac{i \hbar}{2} \right)^n V^{(n)} \left( \frac{x}{2} \right ) \frac{\partial^n }{\partial p^n}  F_{12}^W \left(  \frac{x}{2},p \right) =\\
\\
= N_1  \int d p \hspace{0.2 cm}  e^{i p x / \hbar} \sum_{n=0}^{+ \infty} \frac{1}{n!} \left( \frac{- i \hbar}{2} \right)^n   \left( \frac{i x}{\hbar } \right)^n V^{(n)} \left( \frac{x}{2} \right) F_{12}^W \left(  \frac{x}{2},p \right) = V(x) \psi_1 (x),
\end{array}
\end{equation}

\vspace{0.5 cm}

\begin{equation}
A_3 (x) = N_1 E_1 \int d p \hspace{0.2 cm}  e^{i p x / \hbar} F_{12}^W \left(  \frac{x}{2},p \right) = E_1 \psi_1 (x).
\end{equation}
Assembling all the results, we get:
\begin{equation}
- \frac{\hbar^2}{2m} \frac{d^2 }{d x^2} \psi_1 (x) + V(x) \psi_1 (x) = E_1 \psi_1 (x).
\end{equation}
Similarly, from the second equation in (27), we would get the following eigenvalue equation:
\begin{equation}
- \frac{\hbar^2}{2m} \frac{d^2 }{d x^2} \psi_2 (x) + V(x) \psi_2 (x) = E_2 \psi_2 (x).
\end{equation}
It is easy to prove that the same results would hold even if we had used the alternative formula (55) in the case $\psi_2 (0) =0$.$_{\Box}$

This completes the time independent case.

\subsection{Time dependent pure states}

The time dependent case is slightly more delicate. If we apply naively the formula (54) to a time dependent Wigner function (31), we will get: $\psi_2^* (0,t) \psi_1 (x,t)$. So there is an overall time dependent constant which cannot be distinguished from the own time dependence of the wavefunction $\psi_1 (x,t)$. This is not surprising as the following argument illustrates. Consider a pair of states $| \psi_1 (t) >$, $| \psi_2 (t) >$. From these we can construct the entries of the non diagonal density matrix element $| \psi_1(t)>< \psi_2(t)|$ in the position representation:
\begin{equation}
\rho (x,y) = < x| \psi_1 (t)>< \psi_2 (t)|y>.
\end{equation}
Now suppose we multiply $| \psi_1 (t) >$ and $| \psi_2 (t) >$ by the same time dependent phase $e^{\frac{i}{\hbar} a(t)}$, then the density matrix element (64) remains unchanged. This means that, given some density matrix $\hat{\rho}$, there are several wavefunctions (with different time evolutions) which yield the same physical predictions as does $\hat{\rho}$. From among these we want to obtain the wavefunctions $\psi_1$, $\psi_2$ (unique up to complex multiplicative constants) which obey the Schr\"odinger equation.

The desired functions are given by the following theorem.

\vspace{0.3 cm}
\noindent
{\underline{\bf Theorem 4:}} Suppose that $F_{12}^W (x,p,t)$ is a solution of the Moyal equation and of the pure state constraint (30) and let 
\begin{equation}
\left\{
\begin{array}{l}
\psi_1 (x,t) = \exp \left\{\frac{1}{2} \ln {\cal P}_{12} (x,t) + \frac{i}{\hbar} \int_0^x d x'  < p (x',t)>_{12} - \frac{i}{\hbar} a(t) \right\},\\
\\ 
\psi_2 (x,t) = \exp \left\{\frac{1}{2} \ln {\cal P}^*_{12} (x,t) + \frac{i}{\hbar} \int_0^x d x' < p (x',t)>_{12}^* - \frac{i}{\hbar} a^*(t) \right\},
\end{array}
\right.
\end{equation}
where
\begin{equation}
a(t) = \int_0^t d t' \hspace{0.2 cm} \left[\frac{<p(0,t')>_{12}^2}{2m} + V(0) + {\cal Q}_{12} (0,t') \right] ,
\end{equation}
and ${\cal Q}_{12}$ is the non-diagonal quantum potential:
\begin{equation}
{\cal Q}_{12} (0,t) = - \frac{\hbar^2}{4m} \frac{1}{{\cal P}_{12}} \left[\frac{\partial^2 {\cal P}_{12}}{\partial x^2} - \frac{1}{2 {\cal P}_{12}} \left(\frac{\partial {\cal P}_{12}}{\partial x} \right)^2 \right].
\end{equation}
Then $F_{12}^W (x,p,t)$ can be expressed in terms of $\psi_1$ and $\psi_2$ as the bilinear functional (31) and the functions $\psi_1$, $\psi_2$ obey the Schr\"odinger equation. 

\vspace{0.3 cm}
\noindent
{\underline{\bf Proof:}} If $F_{12}^W$ is a solution of the Moyal equation,
\begin{equation} 
\frac{\partial F_{12}^W}{\partial t} = \left[H, F_{12}^W \right]_M = - \frac{p}{m} \frac{\partial F_{12}^W}{\partial x} + \sum_{n=0}^{+ \infty} \frac{(-1)^n}{(2n+1)!} \left( \frac{\hbar}{2} \right)^{2n} \frac{\partial^{2n+1} V}{\partial x^{2n+1}} \frac{\partial^{2n+1} F_{12}^W}{\partial p^{2n+1}},
\end{equation}
it is then easy to conclude that the dynamics of ${\cal P}_{12}$ and $<p>_{12}$ are dictated by:
\begin{equation}
\left\{
\begin{array}{l}
\frac{\partial {\cal P}_{12}}{\partial t} + \frac{\partial}{\partial x} \left[\frac{{\cal P}_{12}}{m} <p>_{12} \right]=0,\\
\\
\frac{\partial <p>_{12}}{\partial t} = - \frac{1}{{\cal P}_{12}} \frac{\partial}{\partial x} \left( \frac{{\cal P}_{12}}{m} \Delta_{12}^{(2)} p \right) - \frac{\partial}{\partial x} \left( \frac{<p>_{12}^2}{2m}  +   V (x) \right)
\end{array}
\right.
\end{equation}
Since $F_{12}^W$ is a pure state which satisfies eq.(30), then we also have (cf.(52)):
\begin{equation}
\Delta_{12}^{(2)} p(x,t) = - \frac{\hbar^2}{4} \frac{\partial^2}{\partial x^2 } \ln {\cal P}_{12} (x,t).
\end{equation}
It then follows that:
\begin{equation}
\begin{array}{c}
\frac{\partial <p>_{12}}{\partial t} = \frac{\partial}{\partial x} \left\{\frac{\hbar^2}{4m} \left[\frac{1}{{\cal P}_{12}} \frac{\partial^2 {\cal P}_{12}}{\partial x^2} - \frac{1}{2} \left(\frac{1}{{\cal P}_{12}} \frac{\partial {\cal P}_{12}}{\partial x} \right)^2 \right] - \frac{<p>_{12}^2}{2m} - V(x) \right\}=\\
\\
= - \frac{\partial}{\partial x} \left\{ {\cal Q}_{12} (x,t) + \frac{<p(x,t)>_{12}^2}{2m} + V(x) \right\}.
\end{array}
\end{equation}
From these expressions we get:
\begin{equation}
\begin{array}{c}
\dot \psi_1 (x,t) =\left\{\frac{1}{2} \frac{\dot{\cal P}_{12}}{{\cal P}_{12}} + \frac{i}{\hbar} \int_0^x d x' < \dot p (x',t) >_{12} - \frac{i}{\hbar} \dot a (t) \right\} \psi_1 (x,t) =\\
\\
= \left\{- \frac{1}{2m {\cal P}_{12}} \frac{\partial {\cal P}_{12}}{\partial x} <p>_{12} - \frac{1}{2m} \frac{\partial <p>_{12}}{\partial x} - \frac{i}{\hbar} \left[{\cal Q}_{12} (x,t) + \frac{< p(x,t)>_{12}^2}{2m} + V(x) \right]+ \right.\\
\\
\left. + \frac{i}{\hbar} \left[{\cal Q}_{12} (0,t) + \frac{< p(0,t)>_{12}^2}{2m} + V(0) \right] -  \frac{i}{\hbar} \left[{\cal Q}_{12} (0,t) + \frac{< p(0,t)>_{12}^2}{2m} + V(0) \right] \right\} \psi_1 (x,t)=\\
\\
=  \left\{- \frac{<p>_{12}}{2m {\cal P}_{12}} \frac{\partial {\cal P}_{12}}{\partial x}  - \frac{1}{2m} \frac{\partial <p>_{12}}{\partial x} - \frac{i}{\hbar} \left[{\cal Q}_{12} (x,t) + \frac{< p(x,t)>_{12}^2}{2m} + V(x) \right] \right\} \psi_1 (x,t).
\end{array}
\end{equation}
On the other hand we have:
\begin{equation}
\begin{array}{c}
\frac{\partial^2}{\partial x^2} \psi_1 (x,t) = \left\{ \left[\frac{1}{2 {\cal P}_{12}} \frac{\partial {\cal P}_{12}}{\partial x} + \frac{i}{\hbar} <p>_{12} \right]^2 + \frac{1}{2 {\cal P}_{12}} \frac{\partial^2 {\cal P}_{12}}{\partial x^2} - \frac{1}{2} \left( \frac{1}{{\cal P}_{12}} \frac{\partial {\cal P}_{12}}{\partial x} \right)^2 + \frac{i}{\hbar} \frac{\partial <p>_{12}}{\partial x} \right\} \psi_1 (x,t)=\\
\\
= \left\{ - \frac{1}{\hbar^2} <p>_{12}^2 - \frac{2m}{\hbar^2} {\cal Q}_{12} (x,t) + \frac{i}{\hbar} \frac{<p>_{12}}{{\cal P}_{12}} \frac{\partial {\cal P}_{12}}{\partial x} + \frac{i}{\hbar} \frac{\partial <p>_{12}}{\partial x} \right\} \psi_1 (x,t).
\end{array}
\end{equation}
Comparing eqs.(72) and (73), we recover the Schr\"odinger equation (23). A similar result can be obtained for $\psi_2 (x,t)$.

It remains to show that $F_{12}^W$ can be expressed in terms of the functions $\psi_1$, $\psi_2$ (65) according to (31). Since $F_{12}^W$ is a pure state we already know that:
\begin{equation}
F_{12}^W (x,p,t) = \frac{1}{2 \pi} \int dy \hspace{0.2 cm} e^{- i y p} A \left( x - \frac{\hbar y}{2} , t \right) B \left( x + \frac{\hbar y}{2} , t \right),
\end{equation}
where $A$ and $B$ are some complex functions. Substituting in eqs.(65), we get:
\begin{equation}
\left\{
\begin{array}{l}
\psi_1 (x,t) = B (x,t) \exp \left\{\frac{1}{2} \ln \left[\frac{A(0,t)}{B (0,t)} \right] - \frac{i}{\hbar} a(t) \right\}\\
\\
\psi^*_2 (x,t) = A (x,t) \exp \left\{- \frac{1}{2} \ln \left[\frac{A(0,t)}{B (0,t)} \right] + \frac{i}{\hbar} a(t) \right\}
\end{array}
\right.
\end{equation}
We then conclude that:
\begin{equation}
\psi^*_2 \left( x - \frac{\hbar y}{2} , t \right) \psi_1 \left( x + \frac{\hbar y}{2} , t \right) = A \left( x - \frac{\hbar y}{2} , t \right) B \left( x + \frac{\hbar y}{2} , t \right),
\end{equation}
and it follows that $F_{12}^W$ is indeed the bilinear functional of $\psi_1$ and $\psi_2$ (31).$_{\Box}$

Notice that the equations (65) are equally valid for the time independent case. However, the formula (54) is usually easier to evaluate, and it does pay off to use it in the time independent case.

Before we end this section let us make some remarks:\\
\\
1) Equations (30) and (54) were introduced by Tatarskii in ref.\cite{Tatarskii}. He admitted only diagonal state solutions to equation (30). Here we accept non-diagonal pure states as well, because these will be crucial for the construction of mixed states in section 5. Tatarskii considered (54) as the right formula for time dependent states. Here we consider the alternative formula (65) for the time dependent case, because otherwise the wavefunction will have an undefined time dependent phase. Furthermore, we considered a general stargenvalue equation (34) and proved that its solutions are pure states associated to wave functions satisfying the proper eigenvalue equations. The explicit form of these wave functions is given by (54). We checked this explicitly for the energy stargenfunctions (Theorem 3). In summary, the former results provide the complete generalization and specification of Baker's converse construction for pure state Wigner functions. 

Mathematically this specification can be implemented as follows (to make it simpler we consider only real pure states, the extension to the more general case being straightforward): let us denote by ${\cal F}_{pure}^W$ the space of all real, square integrable and normalized functions defined on phase space and which satisfy the condition (30). This space is equivalent to the quantum Hilbert space of distinct physical states. Let $[\psi]$ be the equivalence class of normalized wave functions belonging to the Hilbert space ${\cal H}$ and such that $\phi \in [\psi]$ if and only if there is a global phase factor $e^{i\theta}$ such that $\phi = e^{i\theta} \psi$. Let $[{\cal H}]$ be the set of these equivalence classes. Then there is a one to one map $
\Lambda^W : [{\cal H}] \longrightarrow  {\cal F}_{pure}^W $ defined as:
\begin{equation}
[\psi (x)] \longmapsto \Lambda^W \left( [\psi (x)] \right) = F^W (x,p) = \frac{1}{2 \pi} \int dy \hspace{0.2 cm} e^{- i y p} \psi^* \left( x- \frac{\hbar y}{2} \right) \psi \left( x +  \frac{\hbar y}{2} \right),
\end{equation}
and displaying the following inverse map: $\left( \Lambda^W \right)^{-1} :  {\cal F}_{pure}^W  
\longrightarrow  [{\cal H}]$, with:
\begin{equation}
F^W (x,p) \longmapsto \left( \Lambda^W \right)^{-1} \left(F^W (x,p) \right) =  [\psi (x)]: \, \psi(x)  = N \int dp \hspace{0.2 cm} e^{ipx / \hbar} F^W \left( \frac{x}{2} , p \right),
\end{equation}
where $N$ is a normalization constant (defined up to a global phase factor).
Furthermore, if $\psi(x)$ is an eigenfunction then $F^W(x,p)$ is the corresponding stargenfunction and vice-versa. Finally, 
if $\psi(x,t)$ satisfies the Schr\"odinger equation then $\Lambda^W([\psi(x,t)])$ satisfies the Moyal equation. Conversely, if $F^W(x,p,t) $ satisfies the Moyal equation then there is a wave function $\phi(x,t) \in [\psi(x,t)]$ in the equivalence class of $\psi(x,t)$ (eq.(78)) that satisfies the Schr\"odinger equation. The explicit form of $\phi(x,t)$ is given by eq.(65).\\
\\
2) The pure state quantum condition (30) can be rederived in a different (but equivalent) functional form by using the momentum representation of the original wave functions as a starting point. For a generic quantum state $|\psi>$ the momentum $\tilde{\psi}(p)=<p|\psi>$ and the position $\psi(x)=<x|\psi>$ representations are related by:
\begin{equation}
\tilde{\psi}(p)= \frac{1}{\sqrt{2 \pi \hbar}} \int dx \hspace{0.2cm} \psi(x) e^{-ixp/\hbar} \quad \mbox{and} \quad 
\psi(x)= \frac{1}{\sqrt{2 \pi \hbar}} \int dp \hspace{0.2cm} \tilde{\psi}(p) e^{ixp/\hbar},    
\end{equation}
from where it follows an alternative functional form of the pure state Wigner function:
\begin{equation}
F^W_{pure}(x,p)=\frac{1}{2 \pi} \int dy \hspace{0.2 cm} e^{- i y p} \psi^* \left( x- \frac{\hbar y}{2} \right) \psi \left( x +  \frac{\hbar y}{2} \right)=
\frac{1}{\pi \hbar} \int dk \hspace{0.2 cm} e^{- 2i xk/\hbar} \tilde{\psi}^* \left( p-k \right) \tilde{\psi} \left( p + k  \right).
\end{equation}
We now follow the procedure of eqs.(29,30) and define:
\begin{equation}
\Sigma(y,p)= \int dx \hspace{0.2cm} e^{ixy} F^W(x,p),
\end{equation}
and from (80) we easily conclude that a real, square integrable and normalized phase space function is a pure state Wigner function iff it satisfies:
\begin{equation}
\frac{\partial^2}{\partial y^2} \ln \Sigma(y,p) = \left( \frac{\hbar}{2} \right)^2 \frac{\partial^2}{\partial p^2} \ln \Sigma(y,p) .  
\end{equation}
Hence, the former equation provides an alternative (equivalent) formulation of the pure state quantum condition (30) that in some specific cases might be easier to evaluate than formula (30). Finally, we should also notice that the pure state conditions (29,30) and (81,82) can be rederived for arbitrary non-commuting phase space observables $A, B$ satisfying $[A,B]_M=1$.
\\
\\
3) The pure state quantum condition implies (but it is not equivalent to) Heisenberg's uncertainty principle. In the next section we will prove this result for the more general set of mixed quantum states (which includes as a subset the pure quantum states). Here, we follow Tatarskii through the following elucidative example of the relation between Heisenberg's uncertainty principle and the pure state quantum condition. Consider a Hamiltonian quadratic in the positions and momenta like for instance the simple harmonic oscillator. The Moyal equation reduces to the classical Liouville equation and all references to Planck's constant thus disappear. Indeed the time independent Moyal equation reads: 
\begin{equation}
\left\{
\begin{array}{l}
\left[ H(x,p) , F^W (x,p) \right]_M = - \frac{p}{m} \frac{\partial F^W}{\partial x} + m \omega^2 x \frac{\partial F^W}{\partial p} = 0,\\
\\
H (x,p) = \frac{p^2}{2m} + \frac{1}{2} m \omega^2 x^2.
\end{array}
\right.
\end{equation}
Any function of $H$ solves eq.(83). In particular, let us choose the following solution:
\begin{equation}
F^W (x,p) = \frac{a \omega}{2 \pi} e^{- a H(x,p)},
\end{equation}
where $a$ is a positive real constant. The function $F^W (x,p)$ is real, normalized and square integrable. However if we compute the position and momentum dispersions, we get:
\begin{equation}
\left(\Delta_2 x \right)^2 = \frac{1}{m a \omega^2}, \hspace{1 cm}  \left(\Delta_2 p \right)^2 = \frac{m}{a},
\end{equation}
which means that:
\begin{equation}
\Delta_2 x \cdot \Delta_2 p = \frac{1}{a \omega}.
\end{equation}
Since the constant $a$ is arbitrary, there is no lower bound for the product in eq.(86), in which case Heisenberg's uncertainty principle may be violated. So (84) will in general not be an admissible quantum state.

If, however, we impose the pure state condition (30), we get:
\begin{equation}
a= \frac{2}{\hbar \omega},
\end{equation}
and the product (86) will now respect Heisenberg's bound.
This example also shows that the pure state quantum condition is more restrictive than the Heisenberg uncertainty relations. In fact the Heisenberg bound implies only that $a \le \frac{2}{\hbar \omega}$. We will came back to this issue in the next section.\\ 
\\
4) An interesting question is whether classical dynamics preserves the pure state quantum condition. We will now prove that in general this is not the case. To do so we consider the example of an anharmonic oscillator described by the Hamiltonian:
\begin{equation}
H(x,p)=\frac{p^2}{2m} +\frac{1}{2}mw^2x^2 + \lambda x^4,
\end{equation}
where $\lambda$ is a positive and real coupling constant.
The Moyal equation reads:
\begin{equation}
\frac{\partial F^W}{\partial t} = [H,F^W]_M = -\frac{p}{m} \frac{\partial F^W}{\partial x} + \left(mw^2x+4 \lambda x^3 \right)\frac{\partial F^W}{\partial p}- \lambda \hbar^2 x \frac{\partial^3 F^W}{\partial p^3},
\end{equation}
and up to first order in $t$ we have:
\begin{equation}
F^W_Q(x,p,t)=F_0^W(x,p) + t \left(-\frac{p}{m} \frac{\partial F^W_0}{\partial x} + \left(mw^2x+4 \lambda x^3 \right)\frac{\partial F^W_0}{\partial p}- \lambda \hbar^2 x \frac{\partial^3 F^W_0}{\partial p^3} \right) + {\cal O}(t^2),
\end{equation}
where the subscript $Q$ stands for the {\it quantum} time evolution of $F_0^W(x,p)$. If we use the Poisson bracket instead and calculate the classical (Liouville) time evolution of $F^W_0$ we easily get:
\begin{equation}
F^W_C(x,p,t)=F^W_Q(x,p,t)+ t \lambda \hbar^2 x \frac{\partial^3 F^W_0}{\partial p^3} + {\cal O}(t^2).
\end{equation}
To proceed we calculate $Z(x,j,t)$ up to first order in $t$ both for the classical and quantum time evolutions. We have:
\begin{equation}
\left\{ \begin{array}{lll}
Z_Q(x,j,t)= \int dp \hspace{0.2cm} e^{ijp} F_Q^W(x,p,t) & = & Z_0(x,j)+itZ_1^Q(x,j) + {\cal O}(t^2) \\
Z_C(x,j,t)= \int dp \hspace{0.2cm} e^{ijp} F_C^W(x,p,t) & = & Z_0(x,j)+itZ_1^C(x,j) + {\cal O}(t^2) , 
\end{array} \right.
\end{equation}
where:
\begin{equation}
Z_1^Q(x,j)=\frac{1}{m} \frac{\partial^2 Z_0}{\partial x \partial j}-\left(mw^2xj+4 \lambda x^3j \right) Z_0 - \hbar^2\lambda x j^3 Z_0 = Z_1^C(x,j) -  \hbar^2\lambda x j^3 Z_0(x,j),
\end{equation}
and $Z_0=Z_0(x,j)=\int dp \hspace{0.2cm} e^{ijp} F_0^W(x,p)$. Now notice that:
\begin{equation}
\ln Z_Q(x,j,t)=\ln Z_0(x,j) + \left. \frac{\partial}{\partial t} \ln Z_Q(x,j,t) \right|_{t=0} t+ {\cal O}(t^2)=
\ln Z_0(x,j) +it \frac{Z_1^Q(x,j)}{Z_0(x,j)} + {\cal O}(t^2),
\end{equation}
and thus, up to the first order in $t$, the pure state quantum condition is written:
\begin{equation}
\frac{\partial^2}{\partial j^2} \ln Z_Q(x,j,t)= \left( \frac{\hbar}{2}\right)^2  \frac{\partial^2}{\partial x^2} \ln Z_Q(x,j,t) \Longrightarrow \frac{\partial^2}{\partial j^2} \left(\frac{Z_1^Q(x,j)}{Z_0(x,j)} \right)=\left( \frac{\hbar}{2}\right)^2  
\frac{\partial^2}{\partial x^2} \left(\frac{Z_1^Q(x,j)}{Z_0(x,j)}\right).
\end{equation}
That this equation is satisfied is a consequence of Theorem 1. This result can also be checked explicitly using (93). Here we want to see if eq.(95) is also satisfied if we consider the classical evolution instead. We have:
\begin{eqnarray}
&& \frac{\partial^2}{\partial j^2} \left(\frac{Z_1^C(x,j)}{Z_0(x,j)} \right)  =  \left( \frac{\hbar}{2}\right)^2  
\frac{\partial^2}{\partial x^2} \left(\frac{Z_1^C(x,j)}{Z_0(x,j)}\right)
\Longleftrightarrow  \\
&& \frac{\partial^2}{\partial j^2} \left(\frac{Z_1^Q(x,j)+\hbar^2\lambda x j^3 Z_0(x,j)}{Z_0(x,j)} \right) = \left( \frac{\hbar}{2}\right)^2  
\frac{\partial^2}{\partial x^2} \left(\frac{Z_1^Q(x,j)+\hbar^2\lambda x j^3 Z_0(x,j)}{Z_0(x,j)}\right)
\Longleftrightarrow 6 \hbar^2\lambda x j =0, \nonumber
\end{eqnarray}
where we used eq.(93). The last identity is satisfied only if $\lambda=0$ in which case the anharmonic oscillator reduces to the simple harmonic oscillator for which the Moyal and Liouville time evolutions coincide. We conclude that in general the classical dynamical structure is not compatible with the pure state quantum condition. An interesting question for future research is whether the quantum condition uniquely determines the Moyal dynamics.\\
\\
5) Another interesting result is that the classical measurement procedure does not preserve the pure state quantum condition either. Assume that $F^W(x,p)$ is a classical phase space distribution (i.e. is real, positive defined and normalized). There are in fact few cases where these conditions are compatible with the phase space quantum condition (the problem, as is well known, is positivity). The Gaussian distribution eq.(84,87) is one of those few examples that can be regarded as both a quantum and classical distribution. So assume for the moment that $F^W(x,p)$ is a Gaussian distribution. Under a classical measurement of a generic classical observable $A(x,p)$ with output $a$, $F^W(x,p)$ will collapse to:
\begin{equation}
F^W(x,p) \longrightarrow N F^W(x,p) \delta (A(x,p)-a),
\end{equation}
where $N$ is a normalization constant. The new distribution is still well defined as a classical distribution (is positive, real and normalized) and displays a well known feature of classical measurements: it preserves the original functional form of $F^W(x,p)$ on the phase space hypersurface $A(x,p)=a$. This feature is of course not shared by the quantum measurement and it is trivial to realize that in general the outcoming distribution (97) is not a pure (even distributional) quantum state of the form (17) (take for instance $A(x,p)=p$). In fact, the distribution (97) in general does not satisfy the Heisenberg uncertainty relations (for $A(x,p)=p$ the Heisenberg uncertainty relations would imply that $F^W(x,a)=Const , \, \forall x$) and thus cannot be a pure (distributional) quantum state. We conclude that the preservation of the pure state quantum conditions through the measurement procedure imposes a reformulation of the classical collapse prescription. Again, an interesting question is whether the quantum conditions uniquely determine the standard form of the quantum collapse \cite{nuno4}.

\section{Mixed states}

In this section we generalize the previous results to mixed states. In particular we want to determine when does a phase space function $F^W (x,p)$ represent an admissible mixed quantum state. As always it must be real and normalized. But now we want to find some suitable condition which plays the role of (30).\\
\\
{\underline{\bf Definition:}}
Let $F^W_{pure} (x,p)$ be an arbitrary pure state Wigner function, i.e. a real, square integrable and normalized function which obeys eq.(30). Then $F^W (x,p)$ is an admissible quantum Wigner function if it is real, normalized and satisfies the following "{\it mixed state quantum condition}":
\begin{equation}
\int dx \int dp \hspace{0.2 cm} F^W (x,p) F^W_{pure} (x,p) \ge 0,
\end{equation}
for all pure state Wigner functions $F^W_{pure} (x,p) \in {\cal F}^W_{pure}$.$_{\Box}$ 

We shall now prove the following Lemma:

\vspace{0.3 cm}
\noindent
{\underline{\bf Lemma 3:}} A real and normalized phase space function $F^W (x,p)$ is of the form (21) iff it obeys the "{\it mixed state quantum condition}" (98).

\vspace{0.3 cm}
\noindent
{\underline{\bf Proof:}} Suppose that the real and normalized function $F^W (x,p)$ satisfies the inequality (98). Let us define the candidate density matrix in the position representation:
\begin{equation}
\rho (x,y) = \int dp \hspace{0.2 cm} e^{\frac{ i p}{\hbar} (x-y)} F^W \left( \frac{x+y}{2} , p \right).
\end{equation}
where (at this point) $\rho (x,y)$ might only make sense as a distribution. We have:
\begin{equation}
Tr \hat{\rho} = \int dx \hspace{0.2 cm} \rho (x,x) = \int dx \int dp \hspace{0.2 cm} F^W (x,p) =1,
\end{equation}
\begin{equation}
\rho (x,y)^* = \int dp \hspace{0.2 cm} e^{- \frac{ i p}{\hbar} (x-y)} F^W \left( \frac{x+y}{2} , p \right) = \rho (y,x),
\end{equation}
and so $\hat{\rho}$ is hermitian and normalized. Moreover, for an arbitrary normalized wavefunction $\psi (x)$, we have:
\begin{equation}
\begin{array}{c}
\int du \int dv \hspace{0.2 cm} \psi^* (u) \rho (u,v) \psi (v) = \int dp \int du \int dv \hspace{0.2 cm} e^{- \frac{ i p}{\hbar} (v-u)} F^W \left( \frac{u+v}{2} , p \right) \psi^* (u) \psi (v)=\\
\\
=  \hbar \int dp \int dx \int dy \hspace{0.2 cm} e^{- iyp} F^W \left(x , p \right) \psi^* \left(  x - \frac{\hbar y}{2} \right) \psi \left(  x + \frac{\hbar y}{2} \right) = 2 \pi \hbar \int dx \int dp \hspace{0.2 cm} F^W (x,p) F^W_{pure} (x,p) \ge 0,
\end{array}
\end{equation}
where we used (98) and:
\begin{equation}
F^W_{pure} (x,p) = \frac{1}{2 \pi} \int dy \hspace{0.2 cm} e^{- iyp}  \psi^* \left(  x - \frac{\hbar y}{2} \right) \psi \left(  x + \frac{\hbar y}{2} \right).
\end{equation}
We also performed the substitution:
\begin{equation}
u = x - \frac{\hbar y}{2}, \hspace{1 cm} v = x + \frac{\hbar y}{2} .
\end{equation}
Since $\psi (x)$ is arbitrary, we conclude that the quadratic form associated with the matrix $\hat{\rho}$ is non negative. It follows that $\hat{\rho}$ satisfies all the postulates of the quantum density matrix discussed in section 2. Consequently, it can be expressed as a superposition of pure states as in (10) and it satisfies $Tr \hat{\rho}\hat{\rho} \le 1 \Longrightarrow \int dx dy \hspace{0.1cm} \rho^*(x,y) \rho(x,y) \le 1$. Hence, we can invert the Fourier transform (99) and conclude that $F^W (x,p)$ is of the form (21).

Conversely, if $F^W(x,p)$ satisfies eq.(21) then $F^W(x,p)$ is real, normalized and satisfies: 
\begin{equation}
\int dx \int dp \hspace{0.2 cm} F^W (x,p) F^W_{pure} (x,p)=\sum_{\beta \in B} p_{\beta} \int dx \int dp \hspace{0.2 cm} F^W_{pure,\beta} (x,p) F^W_{pure} (x,p).
\end{equation}
Let $\psi_{\beta}(x),\phi(x)$ be the wave functions associated to $F^{W}_{pure,\beta} (x,p)$ and $F^W_{pure} (x,p)$, respectively. Both $F^{W}_{pure,\beta} (x,p)$ and $ F^W_{pure} (x,p)$ display the standard functional form given by equation (103) and we have:
\begin{equation}
\begin{array}{c}
\int dx \int dp \hspace{0.2 cm} F^{W}_{pure,\beta} (x,p) F^W_{pure} (x,p) = \frac{1}{2 \pi} \int dx \int dy \hspace{0.2 cm} \psi_{\beta}^* \left( x - \frac{\hbar y}{2} \right) \psi_{\beta} \left( x + \frac{\hbar y}{2} \right) \phi^* \left( x - \frac{\hbar y}{2} \right) \phi \left( x + \frac{\hbar y}{2} \right)=\\
\\
= \frac{1}{2 \pi \hbar} \int du \int dv \hspace{0.2 cm} \psi_{\beta}^* \left(u \right) \psi_{\beta} \left( v \right) \phi^* \left( v \right) \phi \left( u \right) = \frac{1}{2 \pi \hbar} \left| \int du  \hspace{0.2 cm} \psi_{\beta}^* \left(u \right)  \phi \left( u \right) \right|^2 \ge 0,
\end{array}
\end{equation}
where we used eq.(104). Substituting this result in eq.(105) we conclude that $F^W(x,p)$ satisfies the mixed state quantum condition (98).$_{\Box}$\\
\\
{\underline{\bf Corollary:}} All pure quantum states obey the mixed state quantum condition.\\
\\
{\underline{\bf Proof:}} This result follows immediately from eq.(106).$_{\Box}$\\
\\
{\underline{\bf Theorem 5:}} The mixed state quantum condition is preserved through the Moyal time evolution.\\
\\
{\underline{\bf Proof:} 
We want to prove that if condition (98) is verified at $t=0$, then it will hold at all later times, provided $F^W (x,p)$ evolves in accordance with Moyal's equation. This a trivial result that follows from theorem 1 and the linearity of the Moyal equation applied to a generic mixed state of the form (21). Alternatively, we shall provide a more direct proof that explicitly exhibits the time evolution of the mixed state quantum condition. Let then:
\begin{equation}
U(t) = e_*^{- \frac{it}{\hbar} H (x,p)},
\end{equation}
be the Weyl map of the quantum evolution operator. The noncommutative exponential is defined by:
\begin{equation}
\left\{
\begin{array}{l}
e_*^{A (x,p)} = \sum_{n=0}^{\infty} \frac{1}{n!} \Omega_n (x,p), \\
\\
\Omega_0 (x,p) \equiv 1, \hspace{1 cm} \Omega_{n+1} (x,p) = A (x,p) * \Omega_n (x,p), \hspace{0.5 cm} n \ge 0 .
\end{array}
\right.
\end{equation}
Then the solution of Moyal's equation (24) is:
\begin{equation}
F^W (x,p,t) = U (t) * F^W (x,p,0) * U (-t).
\end{equation}
Using the identity \cite{Fairlie1}
\begin{equation}
\int dx \int dp \hspace{0.2 cm} A (x,p) * B(x,p) = \int dx \int dp \hspace{0.2 cm} A (x,p)  B(x,p),
\end{equation}
we conclude that:
\begin{equation}
\begin{array}{c}
\int dx \int dp \hspace{0.2 cm} F^W (x,p,t) F^W_{pure} (x,p)= \int dx \int dp \hspace{0.2 cm} U(t)* F^W (x,p,0) * U (-t) * F^W_{pure} (x,p) = \\
\\
= \int dx \int dp \hspace{0.2 cm}  F^W (x,p,0) * U (-t) * F^W_{pure} (x,p) * U(t) = \int dx \int dp \hspace{0.2 cm} F^W (x,p,0) F^W_{pure} (x,p ,-t).
\end{array}
\end{equation}
Since the evolution of a pure state is a pure state, from (98) follows that:
\begin{equation}
\int dx \int dp \hspace{0.2 cm} F^W (x,p,t) F^W_{pure} (x,p) \ge 0, \hspace{0.5 cm} \forall t \in \Re,
\end{equation}
and thus we conclude that $F^W (x,p,t)$ will be of the form (21) at all times.$_{\Box}$

The previous analysis leads to the explicit form of the one to one maps relating the space of physical density matrices acting on the quantum Hilbert space and the space of mixed state Wigner functions. These maps provide a suitable generalization of the ones previously presented in the context of pure state Wigner functions eqs.(77,78). Let us denote by ${\cal F}_{mixed}^W$ the space of real and normalized functions in phase space which obey the mixed state quantum condition (98). Likewise let ${\cal D}({\cal H})$ be the space of linear operators acting on the Hilbert space ${\cal H}$ which obey to the conditions 1) to 3) of section 2. The two spaces are related by the Weyl map:
$V_W : {\cal D} ({\cal H}) \longrightarrow {\cal F}_{mixed}^W $, which is given by:
\begin{equation}
\hat{\rho} (t) \longmapsto \frac{1}{2 \pi \hbar} V_W \left(\hat{\rho} (t) \right) = F^W(x,p,t)  = \frac{1}{2 \pi} \int d y \hspace{0.2 cm} e^{-iyp} < x+ \frac{\hbar y}{2}| \hat{\rho} (t) | x- \frac{\hbar y}{2}>,
\end{equation}
and is one to one. The inverse map is $V_W^{-1} :{\cal F}_{mixed}^W  \longrightarrow {\cal D} ({\cal H})$, and reads:
\begin{equation}
F^W (x,p,t) \longmapsto V_W^{-1} \left( F^W (x,p,t) \right) = \hat{\rho} (t) = \hbar \int dx \int dy \int dp \hspace{0.2 cm} e^{iyp} F^W (x,p,t) |x+ \frac{\hbar y}{2}>< x- \frac{\hbar y}{2}| .
\end{equation}
Obviously, we have $ {\cal F}_{pure}^W  \subset  {\cal F}_{mixed}^W$.

Now that we have completely specified the relation between the kinematical structures of the Wigner and the density matrix formulations and proved their formal equivalence, we can import the notion of complitude from the space ${\cal D}({\cal H})$ to the quantum phase space $ {\cal F}_{mixed}^W$ and subsequently address the related issue of which quasi-distributions appear with which weight in a given mixed state Wigner function (21). 

Suppose that $\left\{|a>, \hspace{0.2 cm} a \in I \right\}$ is a set of orthonormal eigenstates of some operator $\hat A$ which is complete (i.e it generates the full Hilbert space ${\cal H}$). All admissible quantum density matrices are thus expandable in the basis $\left\{|a><b|, \hspace{0.2 cm} a,b \in I \right\}$. These basis elements satisfy the following orthogonality conditions:
\begin{equation}
|a><b|c><d| = \delta_{b,c} |a><d|, \hspace{0.5 cm} \forall a,b,c,d \in I.
\end{equation} 
If we multiply this equation by $\frac{1}{(2 \pi \hbar)^2}$ and apply Weyl's map, we obtain:
\begin{equation}
F_{ab}^W * F_{cd}^W = \frac{1}{2 \pi \hbar} \delta_{b,c} F_{ad}^W, \hspace{0.5 cm} \forall a,b,c,d \in I,
\end{equation}
where $F_{ab}^W (x,p)$ are the non diagonal Wigner functions $\frac{1}{2 \pi \hbar} V_W \left( |a><b| \right)$. They obey the $*$-genvalue equations:
\begin{equation}
A (x,p) * F_{ab}^W (x,p) = a F_{ab}^W (x,p), \hspace{1 cm}  F_{ab}^W (x,p) * A(x,p)  = b F_{ab}^W (x,p),
\end{equation}
where $A(x,p) = V_W \left( \hat A \right)$. Because of the one to one correspondence between ${\cal D}({\cal H})$ and $ {\cal F}_{mixed}^W$, we conclude that $\left\{F_{ab}^W (x,p), \hspace{0.2 cm} a,b \in I \right\}$ generates ${\cal F}_{mixed}^W$. This means that for any $F^W \in {\cal F}_{mixed}^W$, we have:
\begin{equation}
F^W (x,p) = \sum_{a, b \in I} \rho_{ab} F_{ab}^W (x,p), \hspace{0.5 cm}  \rho_{ab} \in {\cal C}.
\end{equation}
From (116,118), we have:
\begin{equation}
F_{nm}^W * F^W * F_{rs}^W = \frac{1}{(2 \pi \hbar)^2} \sum_{a,b \in I}  \rho_{ab} \delta_{m,a} \delta_{b,r} F_{ns}^W = \frac{1}{(2 \pi \hbar)^2} \rho_{mr} F_{ns}^W.
\end{equation}
And so:
\begin{equation}
\rho_{mr} = (2 \pi \hbar)^2 \frac{F_{nm}^W * F^W * F_{rs}^W}{F_{ns}^W}.
\end{equation}
A trivial calculation shows that these are just the elements of the density matrix in the basis $\left\{|a>, \hspace{0.2 cm} a \in I \right\}$:
\begin{equation}
\begin{array}{c}
<r| \hat{\rho}|m> = \int dx \int dy \hspace{0.2 cm} <r|x><y|m> \rho (x,y) = \int dx \int dy  \int dp \hspace{0.2 cm} \psi_r^* (x) \psi_m (y) e^{- \frac{i p}{\hbar} (x-y)} F^W \left( \frac{x+y}{2} , p \right) =\\
\\
= \frac{1}{2 \pi} \sum_{a,b \in I} \rho_{ab} \int dx \int dy \int dz \int dp \hspace{0.2 cm} \psi_r^* (x) \psi_m (y)  e^{- \frac{i p}{\hbar} (x-y) + i z p} \psi_a^* \left( \frac{x+y - \hbar z}{2} \right) \psi_b \left( \frac{x+y + \hbar z}{2} \right) = \\
\\
= \sum_{a, b \in I} \rho_{ab} \delta_{r,b} \delta_{m,a} = \rho_{mr},   
\end{array}
\end{equation}
where $\psi_n (x) = <x|n>$. 

Alternatively we may expand $F^W(x,p)$ in the form $F^W(x,p)=\sum_{\beta \in B} p_{\beta} F^W_{pure,\beta}(x,p)$ where the associated pure states are the solutions of the left and right stargenvalue equations: 
$$
F^W(x,p)*F^W_{pure,\beta}(x,p)=
F^W_{pure,\beta}(x,p)*F^W(x,p)=\frac{p_{\beta}}{2\pi \hbar} F^W_{pure,\beta}(x,p).
$$
This result follows immediately from applying the Weyl map to the diagonal representation of $\hat{\rho}$ (eq.(10)).

To proceed we provide an alternative form of the mixed state quantum condition. 
\\
\\
{\underline{\bf Theorem 6:}} The mixed state quantum condition is equivalent to the condition:
\begin{equation}
<g^*(x,p)*g(x,p)>_{F^W} \equiv \int dx \int dp \hspace{0.2cm} (g^*(x,p)*g(x,p)) F^W(x,p) \ge 0 ,
\end{equation}
to be satisfied for all complex valued functions $g(x,p)$, provided the integral exists.\\
\\
{\underline{\bf Proof:}} We begin by proving that if $F^W(x,p)$ is a pure state then it satisfies eq.(122). 
We have:
\begin{eqnarray}
&& \int dx \int dp \hspace{0.2cm} (g^*(x,p)*g(x,p)) F^W_{pure}(x,p) \nonumber \\
& = & 2\pi \hbar \int dx \int dp \hspace{0.2cm} (g^*(x,p)*g(x,p))* (F^W_{pure}(x,p)*F^W_{pure}(x,p)) \nonumber \\
& = & 2\pi\hbar \int dx \int dp \hspace{0.2cm} (g(x,p)* F^W_{pure}(x,p))^* *(g(x,p)*F^W_{pure}(x,p)) \nonumber \\
& = & \int dx \int dp \hspace{0.2cm} |g(x,p) * F^W(x,p)|^2 \ge 0 ,
\end{eqnarray}  
where we use eq.(110) and the identity (28) for pure states. 
The generalization of the former result for mixed states is straightforward. Using eq.(21) we have:
\begin{equation}
\int dx \int dp \hspace{0.2cm} (g^*(x,p)*g(x,p)) F^W(x,p) = \sum_{\beta \in B} p_{\beta} \int dx \int dp \hspace{0.2cm} (g^*(x,p)*g(x,p)) F^W_{pure,\beta}(x,p) \ge 0,
\end{equation}
where we used the result (123) for each of the pure states $F^W_{pure,\beta}(x,p)$.

To prove the converse result we use eq.(28) and get:
\begin{equation}
\int dx \int dp \hspace{0.2cm} F^W_{pure}(x,p) F^W(x,p)=   2 \pi \hbar \int dx \int dp \hspace{0.2cm} \left([F^W_{pure}(x,p)]^* *F^W_{pure}(x,p)    \right) F^W(x,p) \ge 0   ,
\end{equation}
where we also used the reality of the pure state Wigner function and the condition (122) for $g(x,p)=F^W_{pure}(x,p)$.$_{\Box}$

The condition (122) was introduced in a different context by C. Zachos and T. Curtright and proved to imply the Heisenberg uncertainty relations for arbitrary non-commuting observables \cite{Zachos2}. Hence, we conclude that the mixed state quantum condition implies the Heisenberg uncertainty relations. Finally we prove that:
\\
\\
{\underline{\bf Theorem 7:}} Let ${\cal G}$ be the set of complex value phase space functions $g(x,p)$ that satisfy the normalization condition: $\int dx dp \hspace{0.2cm} g^*(x,p)g(x,p)=1$. Then the set of mixed state Wigner functions is given by:
\begin{equation}
{\cal F}_{mixed}^W=\{g^*(x,p)*g(x,p): \, g(x,p) \in {\cal G} \}.
\end{equation}
\\
{\underline{\bf Proof:}} We begin by proving that all phase space functions of the form $g^*(x,p)*g(x,p)$ with $g(x,p) \in {\cal G}$ are mixed quantum states. In fact $g^*(x,p)*g(x,p)$ is real, normalized and by virtue of theorem 6 it satisfies:
\begin{eqnarray}
&&\int dx \int dp \hspace{0.2cm} F^W(x,p) \left\{ g^*(x,p) * g(x,p) \right\} \ge 0 ,\; \forall F^W \in {\cal F}^W_{mixed} \Longrightarrow \nonumber \\ 
&& \int dx \int dp \hspace{0.2cm} F^W_{pure}(x,p) \left\{ g^*(x,p) * g(x,p) \right\} \ge 0 ,\; \forall F^W_{pure} \in {\cal F}^W_{pure} .
\end{eqnarray}
From Lemma 3 we conclude that $g^*(x,p)*g(x,p)$ is a mixed state.

To prove the converse result let us start by recalling that an arbitrary mixed state can be cast in the form:
\begin{equation}
F^W_{mixed} (x,p)= \sum_{\beta \in B} p_{\beta} F^W_{pure,\beta} (x,p),
\end{equation}
where the pure states satisfy: $F^W_{pure,\beta} (x,p) * F^W_{pure,\alpha} (x,p)=\frac{\delta_{\beta , \alpha}}{2 \pi \hbar} F^W_{pure,\beta} (x,p)$. Let then:
\begin{equation}
g(x,p)= \sum_{\beta \in B} \sqrt{2 \pi \hbar p_{\beta} } F^W_{pure,\beta} (x,p),
\end{equation}
and we have:
\begin{eqnarray}
g^*(x,p) * g(x,p) & = & \sum_{\alpha , \beta \in B} \sqrt{2 \pi \hbar p_{\alpha} }\sqrt{2 \pi \hbar p_{\beta} }      F^W_{pure,\alpha} (x,p) * F^W_{pure,\beta} (x,p) \\ 
& = & \sum_{\alpha , \beta \in B} \sqrt{p_{\alpha} p_{\beta}} \delta_{\alpha,\beta} F^W_{pure,\beta} (x,p) = \sum_{\beta \in B} p_{\beta} F^W_{pure,\beta} (x,p)= F^W_{mixed}(x,p) . \nonumber
\end{eqnarray}
We conclude that an arbitrary mixed state Wigner function (128) can be written as in (126). Finally, notice that the decomposition $g^*(x,p) * g(x,p)$ is defined up to an unitary transformation, i.e. if $g(x,p)$ satisfies (130) then $U(x,p)*g(x,p)$ also satisfies (130) for generic unitary symbol $U(x,p)$.$_{\Box}$

Equation (126) provides an elegant, "covariant" characterization of the space of physical states in quantum phase space. An interesting feature of this equation is that it only depends on the starproduct defined on the phase space. This leads to the conjecture that this characterization (126) might also be valid in the more general setting of curved phase space manifolds \cite{nuno4}.

Some remarks are now in order:

\vspace{0.3 cm}
\noindent
(i) First, notice that the mixed state quantum condition, besides implying the Heisenberg uncertainty relations, also implies the positivity of the marginal distributions. In particular we have:
\begin{equation}
\int dp \hspace{0.2 cm} F^W (x,p) =\sum_{\beta \in B} p_{\beta} \int dp \hspace{0.2 cm} F^W_{pure,\beta} (x,p)   =\sum_{\beta \in B} p_{\beta} \left| \psi^{(\beta)}(x) \right|^2 \ge 0, \, \forall x \in \Re ,
\end{equation}
which (together with the normalization condition) means that we can interpret the marginal distribution as a true probability density for $x$. Likewise, we can show that:
\begin{equation}
 \int dx \hspace{0.2 cm} F^W (x,p) \ge 0 , \, \forall p \in \Re ,
\end{equation}
and thus interpret this quantity as the probability density for $p$.

\vspace{0.3 cm}
\noindent
(ii) The properties of Wigner functions associated with quasi-distributions have been analyzed in the past. In particular, it was shown \cite{Tatarskii} that a real and normalized pure state satisfies:
\begin{equation}
\int d x \int d p \hspace{0.2 cm} \left[F_{pure}^W (x,p) \right]^2 = \frac{1}{2 \pi \hbar}.
\end{equation}
Likewise, for mixed states we have:
\begin{equation}
\int d x \int d p \hspace{0.2 cm} \left[F_{mixed}^W (x,p) \right]^2 \le \frac{1}{2 \pi \hbar}.
\end{equation}
Although these conditions can certainly be regarded as preliminary tests to eliminate spurious states, they are by no means sufficient conditions. The necessary and sufficient conditions for real normalized phase space functions to be pure or mixed quantum states are those stated in equations (30,98). Conditions (133,134) are a corollary of the latter. For instance 
eq.(133) is a straightforward consequence of eq.(28). 

\vspace{0.3 cm}
\noindent
(iii) Another interesting fact is the following. We want to show that if $F^W$ is a pure state and solution of the time-independent Moyal equation, then it must be of the form:
\begin{equation}
F^W (x,p) \propto F_{nn}^W (x,p),
\end{equation}
for some $n$, where $\left\{F_{mm}^W (x,p), \hspace{0.2 cm} m \in I \right\}$ are the diagonal $*$-genstates of the energy:
\begin{equation}
H*F_{mm}^W = F_{mm}^W * H = E_m F_{mm}^W.
\end{equation}
Let then $F^W= \sum_{n,m \in I} \rho_{nm} F_{nm}^W$. If it is a solution of the time-independent Moyal equation, then we have:
\begin{equation}
\sum_{n,m \in I} \rho_{nm} \left( E_n -E_m \right) F_{nm}^W =0.
\end{equation}
Since $\left\{F_{nm}^W (x,p), \hspace{0.2 cm} n, m \in I \right\}$ is a complete basis, we conclude that
\begin{equation}
F^W (x,p) = \sum_{m \in I} \rho_{mm} F_{mm}^W (x,p).
\end{equation}
If in addition we impose that $F^W$ be a pure state, then $\rho_{mm}=0$ except for just one $m \in I$, and we recover (135). This is exactly, what happened in our example in section 4. By imposing the pure state condition, we forced $F^W$ to be a diagonal $*$-genstate (84,87) (in this case the fundamental state of the harmonic oscillator). 
\vspace{0.3 cm}
\noindent

(iv) Finally, it is instructive to return to the example discussed in section 4. There we saw that by imposing the pure state quantum condition we obtained a state that satisfied the Heisenberg relations. In fact these were saturated. Now we may admit states of the form (84) that are mixed. In that case we have to impose the mixed state quantum condition (98). 

Let us then suppose that the space ${\cal F}_{mixed}^W $ is generated by the $*$-genstates of the simple harmonic oscillator. If $F^W$ given by eq.(84) is a mixed state then, in particular, it must satisfy the inequality:
\begin{equation}
\int dx \int dp \hspace{0.2 cm} F^W (x,p) F_{11}^W (x,p) \ge 0,
\end{equation}
where
\begin{equation}
F_{11}^W (x,p) = \frac{1}{\pi \hbar} \left( \frac{4 H}{\omega \hbar} -1 \right)  e^{- \frac{2 H}{\omega \hbar}}.
\end{equation}
The diagonal Wigner function $F_{11}^W$ is the first excited state of the simple harmonic oscillator. It is a solution of the $*$-genvalue equations:
\begin{equation}
H(x,p) * F_{11}^W (x,p ) = F_{11}^W (x,p ) * H(x,p) = E_1 F_{11}^W (x,p ),
\end{equation}
where $E_1 = \frac{3}{2} \omega \hbar$. Upon substitution of (140) in (139) we get:
\begin{equation}
\int dx \int dp \hspace{0.2 cm} F^W (x,p) F_{11}^W (x,p) = \frac{a \omega^2 \hbar}{\pi (2 + a \omega \hbar)^2} \left( \frac{2}{\omega \hbar} -a \right) \ge 0.
\end{equation}
We conclude that:
\begin{equation}
a \le \frac{2}{\omega \hbar},
\end{equation}
and the product (86) automatically satisfies the Heisenberg relations. 

In the case of a pure state the quantum nature of the state is made explicit through the appearance of Planck's constant in the constraint (30). It seems that the mixed state quantum condition (98) is not as explicit. However, if we recall that the pure states appearing in (98) must satisfy (30), then the quantum nature of the mixed state is no longer elusive.

\section{Alternative quasi-distributions}

The Wigner function is the quasi-distribution associated with the Weyl correspondence rule. This means the following. We can use the Wigner function to compute e.g. the mean value of a given operator $\hat A$ (eq.(18)) if we choose to cast $\hat A$ in a fully symmetric form (Weyl order). Alternatively, we may decide to impose that the momenta precede the positions (standard order) or vice-versa (anti-standard order). But in that case, in order to keep the mean value unchanged, we must alter the quasi-distribution. Consider for instance the following exponential: $e^{i \xi \hat q + i \eta \hat p}$. If we expand it, all powers will be completely symmetrized in $\hat q$ and $\hat p$. We conclude that the exponential is written in the Weyl order. Using the Baker-Campbell-Hausdorff formula we may express it in a different form: $e^{i \xi \hat q} e^{i \eta \hat p} e^{\frac{i \hbar}{2} \xi \eta}$. Since the positions precede the momenta, the exponential is now written in the standard order. On the contrary, we may choose to place the momenta before the positions (anti-standard order), in which case we would get: $e^{i \eta \hat p} e^{ i \xi \hat q} e^{- \frac{i \hbar}{2} \xi \eta}$. 

The previous analysis can be generalized in the following way. The exponential can be written in an arbitrary order according to:
\begin{equation}
e^{i \xi \hat q + i \eta \hat p} f (\xi, \eta ),
\end{equation}
where $f (\xi, \eta )$ is called the Cohen function \cite{Lee,Cohen} associated with the ordering prescription in question. We assume it to be an analytic function for all $(\xi, \eta )$. For instance, in the previous examples the Cohen function amounts to $f_S (\xi, \eta ) = e^{- \frac{i \hbar}{2} \xi \eta}$ (standard order), $f_{AS} (\xi, \eta ) = e^{\frac{i \hbar}{2} \xi \eta}$ (anti-standard order) and $f_W (\xi, \eta) =1$ (Weyl order).

For a given correspondence rule (i.e. $f (\xi, \eta)$), the quasi-distribution reads:
\begin{equation}
F^f (x,p,t) = \frac{1}{4 \pi^2} \int d \xi \int d \eta \int d x' \hspace{0.2 cm} <x' + \frac{\eta \hbar}{2} | \hat{\rho} (t)| x' - \frac{\eta \hbar}{2}> f ( \xi , \eta ) e^{i \xi (x' -x) - i \eta p}.
\end{equation} 
We are now faced with the following problem. How can we tell whether a given function belonging to the space of all functions living in a phase-space with non-commutative product $*_f$ and Cohen's function $f (\xi, \eta)$, corresponds to a physical quantum state, i.e. belongs to $ {\cal F}_{mixed}^f$? Moreover, if it is a physical state, is it pure or mixed? And, of course, we would like to reconstruct the wavefunction or the density matrix from the quasi-distribution. A generic quasi-distribution may have very different properties from those of the Wigner function. For instance, the Metha function $F^S (x,p,t)$, associated with the standard order, is not a real function.

To get a feeling of the kind of difficulties one faces, let us consider the "{\it mixed state quantum condition}" (98). If we choose the standard order we would write this condition as:
\begin{equation}
\int dx \int dp \hspace{0.2 cm} F^S (x,p) F_{pure}^{AS} (x,p) \ge 0,
\end{equation}
where $F_{pure}^{AS}$ is the quasi-distribution of an arbitrary pure state in the anti-standard order. So if we are interested in analyzing the properties of states in $ {\cal F}_{mixed}^f$, we are forced to study ${\cal F}_{pure}^{f^{-1}}$ as well. The Wigner function corresponds to the simplest case because $f_W (\xi, \eta)= f_W^{-1} (\xi, \eta) =1$.

The shortest route to solving this problem is the following. From eq.(145), we have the following relation between two quasi-distributions with Cohen's functions $f_1$ and $f_2$:
\begin{equation}
F^1 (x,p,t) = \frac{1}{4 \pi^2} \int dx' \int d p' \int d \xi \int d \eta \hspace{0.2 cm} e^{i \xi (x' -x) + i \eta (p' - p)} \frac{f_1 (\xi , \eta)}{f_2 (\xi , \eta)} F^2 (x' , p', t).
\end{equation}
In particular, if we choose $f_2 = f$ and $f_1 = f_W =1$, we get:
\begin{equation}
F^W (x,p,t) = \frac{1}{4 \pi^2} \int dx' \int d p' \int d \xi \int d \eta \hspace{0.2 cm} e^{i \xi (x' -x) + i \eta (p' - p)} f^{-1} (\xi , \eta) F^f (x' , p', t).
\end{equation}
The momentum generating function (29) then reads:
\begin{equation}
Z (x,j,t) = \int dp \hspace{0.2 cm} e^{ijp} F^W (x,p,t) = \frac{1}{2 \pi} \int dx' \int d p' \int d \xi \hspace{0.2 cm} e^{i \xi (x' -x) + i j p'} f^{-1} (\xi , j) F^f (x' , p', t). 
\end{equation}
Having calculated $F^W$ and $Z$ the whole analysis described in sections 4 and 5 can be directly applied. All the formulae presented depend only on $Z$ and $F^W$. We can then determine whether $F^f$ is a physical state, whether it is pure or mixed and we can construct the wavefunction or the density matrix from it.

In the remainder of this section we will just state the explicit formulae for the most popular quasi-distributions found in the literature.

For the Metha function $\left( f_S (\xi, \eta) = e^{- \frac{i \hbar}{2} \xi \eta} \right)$, we have:
\begin{equation}
\left\{
\begin{array}{l}
F^W (x,p,t) = \frac{1}{\pi \hbar} \int d x' \int d p' \hspace{0.2 cm} e^{- \frac{2i}{\hbar} (x' -x) (p'-p)} F^S (x',p',t),\\
\\
Z (x,j,t) = \int dp \hspace{0.2 cm} e^{- i j p} F^S \left( x + \frac{\hbar j}{2} , p ,t \right).
\end{array}
\right.
\end{equation}
Similarly, for the Kirkwood function $\left( f_{AS} (\xi, \eta) = e^{\frac{i \hbar}{2} \xi \eta} \right)$, $F^W$ and $Z$ read:
\begin{equation}
\left\{
\begin{array}{l}
F^W (x,p,t) = \frac{1}{\pi \hbar} \int d x' \int d p' \hspace{0.2 cm} e^{ \frac{2i}{\hbar} (x' -x) (p'-p)} F^{AS} (x',p',t),\\
\\
Z (x,j,t) = \int dp \hspace{0.2 cm} e^{i j p} F^{AS} \left( x + \frac{\hbar j}{2} , p ,t \right).
\end{array}
\right.
\end{equation}
The $P$-function of Glauber and Sudarshan is associated with normal ordering. Let us define the creation and annihilation operators of the simple harmonic oscillator:
\begin{equation}
\hat a = \frac{1}{\sqrt{2 \hbar m \omega}} \left( m \omega \hat q + i \hat p \right), \hspace{1 cm} \hat a^{\dagger} = \frac{1}{\sqrt{2 \hbar m \omega}} \left( m \omega \hat q - i \hat p \right).
\end{equation}
Normal ordering consists in writing the creation operators $\hat a^{\dagger}$ before the annihilation operators $\hat a$. The associated Cohen function is $f_N ( \xi, \eta) = e^{\frac{\hbar \xi^2}{4 m \omega} + \frac{\hbar m \omega \eta^2}{4}}$. We then get:
\begin{equation}
\left\{
\begin{array}{l}
F^W (x,p,t) = \frac{1}{\pi \hbar} \int d x' \int d p' \hspace{0.2 cm} e^{- \frac{m \omega}{\hbar} (x' -x)^2 - \frac{1}{\hbar m \omega}  (p'-p)^2 } F^N (x',p',t),\\
\\
Z (x,j,t) = \sqrt{\frac{m \omega}{\pi \hbar}} e^{- \frac{\hbar m \omega }{4} j^2}  \int d x' \int dp' \hspace{0.2 cm} e^{- \frac{m \omega}{\hbar} (x'-x)^2 + i j p' } F^N \left( x' , p' ,t \right).
\end{array}
\right.
\end{equation}
The Husimi function is defined by $f_H ( \xi, \eta) = e^{- \frac{\hbar \xi^2}{4 m \kappa} - \frac{\hbar m \kappa \eta^2}{4}}$, where $\kappa$ is some positive real constant with dimensions of a frequency. It then follows:
\begin{equation}
\left\{
\begin{array}{l}
F^W (x,p,t) = \frac{1}{4 \pi^2} \int d x' \int d p' \int d \xi \int d \eta \hspace{0.2 cm} e^{ \frac{\hbar \xi^2}{4 m \kappa} + i \xi (x'-x)} e^{ \frac{\hbar m \kappa \eta^2}{4} + i \eta (p'-p)}  F^H (x',p',t),\\
\\
Z (x,j,t) = \frac{1}{2 \pi} e^{ \frac{\hbar m \kappa}{4} j^2}  \int d x' \int dp' \int d \xi \hspace{0.2 cm} e^{\frac{\hbar \xi^2}{4 m \kappa}  + i \xi (x'-x) + i j p'} F^H \left( x' , p' ,t \right).
\end{array}
\right.
\end{equation}
In particular for $\kappa = \omega$, we recover the $Q$-function $F^{AN} (x,p,t)$ associated with anti-normal ordering.

\section{Example 1: the simple harmonic oscillator}

To illustrate the results of section 4, let us consider the time independent harmonic oscillator. For simplicity, we choose units such that $\hbar =1$ and take $m = \omega =1$. The Hamiltonian reads:
\begin{equation}
H (x,p) = \frac{p^2}{2} + \frac{x^2}{2}.
\end{equation}
In refs.\cite{Groenewold,Flato,Fairlie1} it was shown that the normalized diagonal $*$-genfunctions are given by:
\begin{equation}
F_{nn}^W (x,p) = 2 (-1)^n e^{-2 H} L_n (4 H), \hspace{0.5 cm} n=0,1,2, \cdots,
\end{equation}
where $L_n (x)$ are the Laguerre polynomials. They are solutions of the $*$-genvalue equations:
\begin{equation}
H (x,p) * F_{nn}^W (x,p) = F_{nn}^W (x,p) * H(x,p) = E_n F_{nn}^W (x,p),
\end{equation}
with
\begin{equation}
E_n = n + \frac{1}{2} , \hspace{0.5 cm} n = 0,1,2,  \cdots
\end{equation}
It was also shown that the Laguerre polynomials admit the following integral representation \cite{Flato}:
\begin{equation}
L_n (x) = \frac{(-1)^n}{2 \pi i} \oint_C dz \hspace{0.2 cm} (z^2 +1)^{-1} z^{-2n-1} \exp \left( \frac{x z^2}{z^2 +1} \right), \hspace{0.5 cm} n=0,1,2, \cdots,
\end{equation}
where $C$ is a closed contour such that $z=0$ is the only singularity in its interior. Substituting (156,159) in (54) we get:
\begin{equation}
\begin{array}{c}
\psi_n (x) = N_n \int dp \hspace{0.2 cm}e^{ipx} F_{nn}^W \left( \frac{x}{2} , p \right)= \frac{N_n}{\pi i} \oint_C dz \hspace{0.2 cm} (z^2 +1)^{-1} z^{-2n-1} \times\\
\\
\times \int dp \hspace{0.2 cm} e^{ipx} \exp \left[ - \left( \frac{1- z^2}{1+ z^2} \right) \left( p^2 + \frac{x^2}{4} \right) \right]= \frac{N_n}{i \sqrt{\pi}} \oint_C dz \hspace{0.2 cm} \frac{z^{-2n-1}}{\sqrt{1 -z^4}} e^{- \frac{x^2}{2} \left( \frac{1+ z^4}{1 - z^4} \right) }. 
\end{array}
\end{equation}
If $n$ is odd we get $(n=2m+1)$:
\begin{equation}
\psi_{2m+1} (x) = \frac{N_{2 m +1}}{i \sqrt{\pi}} e^{- \frac{x^2}{2}} \oint_C dz \hspace{0.2 cm} \frac{z^{-4m -3}}{\sqrt{1 -z^4}} \exp \left[ \frac{x^2 z^4}{z^4 -1} \right].
\end{equation}
Applying Cauchy's residue theorem, we conclude that the contour integral in the previous equation vanishes. Consequently, the odd modes will have to be evaluated according to the alternative formula (55). 

Let us first compute the even terms $(n=2m)$:
\begin{equation}
\psi_{2m} (x) = \frac{N_{2 m}}{i \sqrt{\pi}} \oint_C dz \hspace{0.2 cm} \frac{z^{-4m -1}}{\sqrt{1 -z^4}} \exp \left[- \frac{x^2}{2} \left( \frac{1+ z^4}{1- z^4 } \right) \right] = \frac{N_{2 m}}{i \sqrt{\pi}} e^{- \frac{x^2}{2}} C_{2m} (x) ,
\end{equation}
where
\begin{equation}
C_{2m} (x) =  \oint_C dz \hspace{0.2 cm} \frac{z^{-4m-1}}{\sqrt{1 -z^4}} \exp \left[\frac{x^2 z^4}{z^4-1} \right].
\end{equation}
If we expand $C_{2m} (x)$ in powers of $x$ using Cauchy's theorem, we get.
\begin{equation}
C_{2m} (x) = \frac{2 \pi i (-1)^m}{m! 2^{2m}} \left\{2^{2m} x^{2m} + \sum_{k=1}^m (-1)^k 2^{2m -k} x^{2(m-k)} \left( 
\begin{array}{c}
2m\\
2k
\end{array}
\right) \cdot 1 \cdot 3 \cdot 5 \cdots (2k-1) \right\}.
\end{equation}
The term in curly brackets is readily identified as the Hermite polynomial of order $2m$ \cite{Gradshteyn}:
\begin{equation}
C_{2m} (x) = \frac{\pi i (-1)^m}{m! 2^{2m-1}} H_{2m} (x).
\end{equation}
Consequently, equation (162) reads:
\begin{equation}
\psi_{2m} (x) = N_{2m} \frac{\sqrt{\pi} (-1)^m}{m! 2^{2m-1}} e^{- \frac{x^2}{2}} H_{2m} (x) = \frac{(-1)^m}{2^m \left[\sqrt{\pi} (2m)! \right]^{1/2}} h_{2m} (x),
\end{equation}
where the constant $N_{2m}$ was determined by requiring that $\psi_{2m}$ be normalized and $h_n (x)$ is the hermite function: $h_n (x) = (-1)^n e^{- x^2 /2}  H_n (x)=e^{x^2 /2} \frac{d^n}{d x^n} e^{- x^2}$.

It is interesting to note that, from this procedure, we obtain an integral representation for the even Hermite polynomials:
\begin{equation}
H_{2m} (x) = m! 4^m (-1)^m \oint_C \frac{dz}{2 \pi i} \hspace{0.2 cm} \frac{z^{-4m-1}}{\sqrt{1-z^4}} \exp \left( \frac{x^2 z^4}{z^4 -1} \right),
\end{equation}
and also a relation between Laguerre and Hermite polynomials:
\begin{equation}
\int dy \hspace{0.2 cm} e^{(x+iy)^2} L_{2m} \left(2(x^2 +y^2) \right) = \frac{\sqrt{\pi} (-1)^m}{m! 4^m} H_{2m} (2x).
\end{equation}
Let us now try to compute the odd solutions:
\begin{equation}
\begin{array}{c}
\psi_{2m+1} (x) = N_{2m+1}' \int dp \hspace{0.2 cm} e^{i p x} \frac{\partial}{\partial x} F_{2m+1, 2m+1}^W \left( \frac{x}{2} , p \right)\\
\\
= \frac{x N_{2m+1}'}{2 i \sqrt{\pi}} \oint_C dz \hspace{0.2 cm} z^{-4m-3} \left( \frac{z^2-1}{z^2 +1} \right)  \left( 1-z^4 \right)^{- 1/2} \exp \left[ - \frac{x^2}{2} \left( \frac{1+z^4}{1- z^4} \right) \right] \\
\\
= \frac{x N_{2m+1}' }{i \sqrt{\pi}} \oint_C dz \hspace{0.2 cm} z^{-4m-1}  \left( 1-z^4 \right)^{- 3/2} \exp \left[ - \frac{x^2}{2} \left( \frac{1+z^4}{1- z^4} \right) \right] \\
\\
- \frac{x N_{2m+1}'}{2 i \sqrt{\pi}} \oint_C dz \hspace{0.2 cm} z^{-4m-3}  \left( 1-z^4 \right)^{- 3/2} \left(1+z^4 \right) \exp \left[ - \frac{x^2}{2} \left( \frac{1+z^4}{1- z^4} \right) \right].
\end{array}
\end{equation}
A simple calculation shows that the second integral in the previous expression vanishes. On the other hand we can easily show that:
\begin{equation}
\frac{d}{dx} \psi_{2m} (x) - x \psi_{2m} (x) = - \frac{2x N_{2m}}{i \pi} \oint_C dz \hspace{0.2 cm} z^{-4m-1}  \left( 1-z^4 \right)^{- 3/2} \exp \left[ - \frac{x^2}{2} \left( \frac{1+z^4}{1- z^4} \right) \right] .
\end{equation}
Comparing (169) and (170), we conclude that:
\begin{equation}
\psi_{2m+1} (x) = \frac{\pi (-1)^{m+1} N_{2m+1}'}{m! 4^m} \left[\frac{d}{dx} h_{2m} (x) - x h_{2m} (x) \right].
\end{equation}
It is well known that the Hermite functions satisfy the equations \cite{Gradshteyn}: $h_n' = x h_n + h_{n+1}$. Consequently, we get the following normalized solution:
\begin{equation}
\psi_{2m+1} (x) = \frac{(-1)^{m+1}}{2^m \left[2 \sqrt{\pi} (2m+1)! \right]^{1/2}} h_{2m+1} (x).
\end{equation}
Moreover, we also get an integral representation for the odd Hermite polynomials:
\begin{equation}
H_{2m+1} (x) = 2x 4^m m! (-1)^m \oint_C \frac{dz}{2 \pi i}  \hspace{0.2 cm} z^{-4m-1}  \left( 1-z^4 \right)^{- 3/2} \exp \left( \frac{x^2 z^4}{z^4 -1 } \right).
\end{equation}
From (167,173) we can construct an unified expression for the Hermite polynomials $H_n (x)$ ($n=0,1,2, \cdots$):
\begin{equation}
H_n (x) = \left(\frac{x}{i} \right)^{\frac{1 + (-1)^{n+1}}{2}} (2i)^n \left[ \left(\frac{2n + (-1)^n -1}{4} \right)! \right]  \oint_C \frac{dz}{2 \pi i}  \hspace{0.2 cm} z^{-2n + (-1)^{n+1} }  \left( 1-z^4 \right)^{\frac{(-1)^n}{2} -1} e^{\frac{x^2 z^4}{z^4 -1 }}. 
\end{equation}

\section{Example 2: the free Gaussian wave packet}

Let us consider the following real Gaussian Wigner function at time $t=0$
\begin{equation}
F^W (x,p,0) = \frac{1}{\pi \hbar} \exp \left[- \frac{x^2}{2 \sigma_0^2} - \frac{2 \sigma_0^2}{\hbar^2} (p-p_0)^2 \right].
\end{equation}
Now suppose that it evolves according to the free Moyal equation $(V(x)=0)$:
\begin{equation}
\frac{\partial F^W}{\partial t} = \left[\frac{p^2}{2m} , F^W \right]_M = - \frac{p}{m} \frac{\partial F^W}{\partial x}.
\end{equation}
The solution is:
\begin{equation}
F^W (x,p,t) = F^W \left(x - \frac{p}{m}t, p,0 \right) = \frac{1}{\pi \hbar} \exp \left[- \frac{1}{2 \sigma_0^2} \left( x - \frac{p}{m}t \right)^2 - \frac{2 \sigma_0^2}{\hbar^2} (p-p_0)^2 \right].
\end{equation}
A straightforward calculation shows that:
\begin{equation}
\left\{
\begin{array}{l}
{\cal P} (x,t) = \left( 2 \pi \sigma_t^2 \right)^{- 1/2} \exp \left[ - \frac{(x-ut)^2}{2 \sigma_t^2} \right],\\
\\
<p(x,t) > = p_0 + \frac{\hbar^2 t (x- ut)}{4m \sigma_0^2 \sigma_t^2},\\
\\
{\cal Q} (x,t) = \frac{\hbar^2}{4m} \left[\frac{1}{\sigma_t^2} - \frac{(x-ut)^2}{2 \sigma_t^4} \right],
\end{array}
\right.
\end{equation}
where
\begin{equation}
u = \frac{p_0}{m}, \hspace{0.5 cm} \sigma_t = \sigma_0 \sqrt{1+ \frac{\hbar^2 t^2}{4m^2 \sigma_0^4}}.
\end{equation}
If we write $\psi$ in the polar form, $\psi = Re^{i S /\hbar}$ (where $S$ and $R$ are real functions), we get from (65,177,178):
\begin{equation}
R(x,t) = \left[ {\cal P} (x,t) \right]^{1/2} = \left( 2 \pi \sigma_t^2 \right)^{- 1/4} \exp \left[ - \frac{(x-ut)^2}{4 \sigma_t^2} \right],
\end{equation}
and
\begin{equation}
\begin{array}{c}
S(x,t) = \int_0^x d x' <p (x',t)> - a(t) = \\
\\
=p_0 x + \frac{\hbar^2 t x}{8m \sigma_0^2 \sigma_t^2} (x-2 u t) - \int_0^t dt' \hspace{0.2 cm} \left\{\frac{\hbar^2}{2m} \left[\frac{1}{2 \sigma_{t'}^2} - \frac{u^2 t'^2}{4 \sigma_{t'}^4} \right] + \frac{1}{2m} \left[p_0 - \frac{\hbar^2 t'^2 u}{4m \sigma_0^2 \sigma_{t'}^2} \right]^2 \right\}.
\end{array}
\end{equation}
Using the fact that $\dot{\sigma}_t = \frac{\hbar^2 t}{4 m^2 \sigma_0^2 \sigma_t}$, we get:
\begin{equation}
S(x,t) = -\frac{\hbar}{2} \arctan \left( \frac{\hbar t}{2 m \sigma_0^2} \right) + p_0 \left( x - \frac{1}{2} u t \right) + \frac{\hbar^2 t}{8m \sigma_0^2 \sigma_t^2} (x-ut)^2.
\end{equation}
From (180,182) we finally get:
\begin{equation}
\psi (x,t) = \left( 2\pi s_t^2 \right)^{- 1/4} \exp \left[ - \frac{(x-ut)^2}{4 \sigma_0 s_t} + \frac{i}{\hbar} p_0 \left( x - \frac{1}{2} u t \right) \right],
\end{equation}
where
\begin{equation} 
s_t = \sigma_0 \left( 1 + \frac{i \hbar t}{2m \sigma_0^2} \right).
\end{equation}
This is well known to be the time evolution of the free Gaussian wave packet \cite{Holland}.

\section{Conclusions}

Let us briefly summarize the results of this paper and comment on some possible future research: we determined necessary and sufficient conditions for pure and mixed phase space quantum states. We showed that this kinematical structure is fully compatible with Wigner quantum mechanics and used the emerging formalism to completely rederive standard operator quantum mechanics from the phase space formulation. These results extend and fully specify Baker's converse construction both to a general stargenvalue equation (associated to a generic phase space symbol) and to the time dependent case. 

Furthermore, we proved that the quantum conditions imply the Heisenberg uncertainty relations, a correction to the classical Liouville equation and to the classical measurement procedure. 
These results lead to the speculation of whether the quantum conditions can acquire the status of a first principle determining the key features of quantum mechanics, i.e. besides the Heisenberg uncertainty relations also fully determining the non-local Moyal dynamics (i.e. the interfering trajectories) and the standard form of quantum collapse (i.e. the interference between the observer and the system during the measurement process). This will be the subject of a future work \cite{nuno4}.

Finally, we determined the explicit form of the quantum conditions for several different quantum phase space ordering prescriptions. For all the studied cases the boundary conditions for the quantum pure state differential equation are completely specified by the values of ${\cal P}_{12}(x)$ and $<p(x)>_{12}$ (eqs.(49,50)). These two quantities also play an important role in the De Broglie-Bohm formulation of quantum mechanics \cite{Holland}. In fact they also uniquely determine the pure state De Broglie-Bohm quasi-distribution. This property together with the fact that De Broglie-Bohm theory can be formulated in phase space in terms of Cohen's quasi-distribution formalism \cite{Cohen,nuno1,nuno9} leads to the conjecture that the results of this paper can be properly extended to the De Broglie-Bohm representation. Hence, an interesting issue for future research would be to determine the explicit form of the pure state quantum condition in the De Broglie-Bohm formulation and show (if possible) that its solutions display the well known functional form of the pure state De Broglie-Bohm quasi-distributions.

\vspace{1 cm}

\begin{center}

{\large{{\bf Acknowledgments}}} 

\end{center}

\vspace{0.3 cm}
\noindent
We would like to thank Aleksandar Mikovic and Ant\'onio Bivar Weinholtz for useful insights.
This work was partially supported by the grants POCTI/MAT/45306/2002 and POCTI/FNU/49543/2002.

\end{document}